\documentclass[preprints,review,accept,oneauthor,pdftex]{mdpi}
\usepackage{graphicx}
\firstpage{1} 
\pubvolume{0}
\issuenum{0}
\articlenumber{0}
\pubyear{2021}
\copyrightyear{2021}
\datereceived{} 
\dateaccepted{} 
\datepublished{} 
\hreflink{https://doi.org/} 

\Title{A short review on clustering dark energy}

\TitleCitation{Title}

\Author{Ronaldo C. Batista \orcidA{0000-0002-7655-8719}}

\AuthorNames{Ronaldo C. Batista}

\AuthorCitation{Batista, R. C.}

\address{%
\quad Escola de Ci\^encias e Tecnologia\\
Universidade Federal do Rio Grande do Norte\\ 
Natal, RN, Brazil; rbatista@ect.ufrn.br\\
}

\corres{Correspondence: rbatista@ect.ufrn.br}

\Title{A short review on clustering dark energy}

\abstract{We review dark energy models which can present non-negligible fluctuations
on scales smaller than Hubble radius. Both linear and nonlinear evolutions
of dark energy fluctuations are discussed. The linear evolution has
a well-established framework, based on linear perturbation theory
in General Relativity, and is well studied and implemented in numerical
codes. We highlight the main results from linear theory to explain
how dark energy perturbations become important on the scales of interest for
structure formation. Next, we review some attempts to understand the
impact of clustering dark energy models in the nonlinear regime, usually based
on generalizations of the Spherical Collapse Model. We critically
discuss the proposed generalizations of the Spherical Collapse Model
that can treat clustering dark energy models and their shortcomings. Proposed
implementations of clustering dark energy models in halo mass functions
are reviewed. We also discuss some recent numerical simulations capable
of treating dark energy fluctuations. Finally, we make an overview of the observational
predictions based on these models.}


\begin{document}

\section{Introduction}

Since the discovery of the accelerated expansion of the Universe \citet{Riess:1998cb,Perlmutter:1998np},
a great variety of explanations have been proposed. The most simple
and well-studied proposal is that the accelerated expansion is caused
by the Cosmological Constant, $\Lambda$, which is constant in space
and time and naturally possesses no fluctuations. In this model, Dark Energy (DE)
only modifies the background cosmological evolution, then the modifications
in linear and nonlinear evolution of cosmological perturbations are
straightforward to implement. Together with the Cold Dark Matter (CDM),
which accounts for roughly $25\%$ of the Universe energy density,
the $\Lambda$CDM model provides an outstanding description of cosmological
data obtained so far, e.g., \citet{Aghanim2020,Abbott2021,Asgari2021}

Although $\Lambda$CDM is very successful in describing almost all
cosmological observations, on theoretical grounds, it is challenged
by the Cosmological Constant Problem \citet{Weinberg1989a,Carroll:2000fy}
and the Cosmic Coincidence Problem \citet{Zlatev:1998tr}. More recently,
direct measurements of the Hubble constant, $H_{0}$, have also been
challenging the $\Lambda$CDM model, showing a disagreement of about
$5\sigma$ with respect to the inferred value from Cosmic Microwave
Background (CMB) data \citet{Verde2019,DiValentino2021b}. A less
significant tension, about $3\sigma$, in the $\Lambda$CDM model
is related to the predicted normalization of the matter power spectrum,
$\sigma_{8}$, the $S_{8}=\sigma_{8}\sqrt{\Omega_{m0}/0.3}$ parameter \citet{Perivolaropoulos2021,DiValentino2021}.
A recent analysis of possible solutions can be found in \citet{Schoeneberg2021}.

Given these difficulties with $\Lambda$, a profusion of alternative
models to explain the cosmic acceleration were proposed. One of the
first and most popular alternatives to $\Lambda$ is the quintessence
class of models. In these models, a new scalar field minimally coupled
to gravity and with no direct interactions to other types of matter
plays the role of DE. This kind of model was studied even before the
discovery of accelerated expansion by \citet{Peebles:1987ek,Wetterich:1987fm}.
In quintessence models, DE is time-dependent and its Equation of State
(EoS) parameter track the background, possibly alleviating the Coincidence
Problem \citet{Caldwell:1997ii,Zlatev:1998tr,Steinhardt:1999nw}.

Since the quintessence field is dynamical, it necessarily has fluctuations.
These fluctuations, however, are relevant only on scales of the order
of Hubble radius \citet{Ma:1999dwa,Brax:2000yb,DeDeo:2003te}. On
the small scales of interest for structure formation, quintessence
perturbations are much smaller than matter (dark matter plus baryons)
perturbations and thus are usually neglected. This tiny amount of
DE perturbations on small scales is a consequence of the sound speed
of the field perturbations, which has a constant value $c_{s}=1$.
This value of the sound speed implies that the sound-horizon scale
of quintessence, a scale below which the perturbations are strongly
suppressed by pressure support, is of the order of Hubble radius,
$c_{s}/H_{0}$.

One of the first proposed models beyond the quintessence that can
possible present relevant perturbations on small scales is the tachyon
scalar field \citet{Sen:2002an,Padmanabhan:2002cp,Bagla:2002yn}.
The main difference with respect to quintessence is that the tachyon
field has a time-varying speed of sound, which can be smaller than
unity, thus allowing the field to cluster more effectively on smaller
scales.

It was also observed that even more general scalar field models could
be constructed. The so-called k-essence models were initially proposed
in the context of Inflation \citet{ArmendarizPicon1999,Garriga:1999vw}.
In such models, one has the freedom to choose both the kinetic term
and the potential of the scalar field, which translates into liberty
to choose the EoS parameter and $c_{s}$. Therefore, in this class
of models, DE can have an arbitrarily low speed of sound, thus its
perturbations can be the same order of magnitude of matter perturbations.
In this scenario, DE has the potential to impact structure formation
beyond the background level.

More recently, the class of Horndesky theories, \citet{Horndeski:1974wa},
was rediscovered and it was shown that both quintessence and k-essence
models are sub-classes of Horndesky. At the linear perturbation level,
these sub-classes are parameterized by the $\alpha_{K}$ parameter
\citet{Bellini2014}. In the context of Horndesky theories, many models
beyond k-essence exist, including non-minimally coupled scalar fields,
which also modify gravitational interaction. The various types of
models that can explain the accelerated expansion can also be described
in the Effective Field Theory framework \citet{Gubitosi2013}. Many
of these proposals are discussed in \citet{Amendola2018}.

In this review, we will focus on DE models described by minimally
coupled to gravity scalar fields. We also restrict our attention to
fields with no direct interaction with other matter fields. The k-essence
class of models can be parametrized as perfect fluids defined by two
time-dependent functions, $w\left(t\right)$ and $c_{s}\left(t\right)$
and this description will suffice to analyze how large DE fluctuations
can be and estimate their observational impact.

The first step to understand DE fluctuations is to study them at the linear perturbative level, which is described by the well-established theory of linear cosmological perturbations, e.g., \citet{Kodama:1985bj,Mukhanov:1990me,Ma:1995ey}.
In this context, the study of DE perturbations is straightforward,
but limited to large scales that did not develop nonlinear matter
fluctuations. Difficulties arise when trying to study DE fluctuations
in the nonlinear regime. Historically, structure formation was studied
using the Newtonian theory, which can not deal
with relativistic fluids, such as DE. The obvious approach of using
full-blown General Relativity to include DE fluctuations in structure
formation studies is undoubtedly challenging. In fact, relativistic
studies of structure formation with Clustering Dark Energy (CDE) were developed quite recently \citet{Dakin2019,Hassani2019,Hassani2020}.

The first attempt to study Clustering DE (CDE) models in the nonlinear
regime used an integration between the Newtonian Spherical Collapse
Model (SCM) with a ``local'' Klein-Gordon equation, modified to
permit the clustering of quintessence \citet{Mota:2004pa}. This study
was able to show important new effects due to DE fluctuations, which
were later confirmed by more general and well-justified models. For
instance, depending on the evolution of its EoS, DE fluctuations can
become nonlinear, impact the nonlinear evolution of matter fluctuations
and change the virialization state of matter halos. Hence, it became
clear that CDE can impact structure formation. Later on, some authors
constructed a more formal and general framework to include CDE in
the SCM, \citet{Abramo2007,Creminelli2010,Basse2011}.

The main focus of this review is to describe and discuss the applicability
of the generalizations of the SCM capable of treating CDE models.
We also pay special attention to the corresponding modifications on
the Halo Mass Functions (HMF), which, up to now, have not been explored
by numerical simulations. Moreover, we review the impact of CDE on
cosmological observables and prospects for its detection.

The plan for this review is the following. Section 2 presents the
essentials of relativistic perturbation theory that describe DE perturbations
and their scale dependence. In Section 3, we review quintessence and
k-essence models, highlighting the conditions under which relevant
DE perturbations can be present. Section 4 discusses the SCM
and its generalizations to study homogeneous and inhomogeneous DE
models. Section 5 presents the Pseudo-Newtonian description
of the SCM, which permits direct contact with linear relativistic
perturbations on small scales and provides a clear and general framework
to study CDE models in the nonlinear regime. We discuss the impact
of DE fluctuations on the critical density threshold and HMF in Sections 6 and 7, respectively. The observational impact of
CDE is reviewed in Section 8. We discuss the main results and perspectives
in Section 9.

\section{Linear perturbations}

To understand the basic behaviour of DE perturbations, it is enough
to consider scalar perturbations in the absence of anisotropic stresses.
In this case, the perturbed line element in the Newtonian gauge is given by
\citet{Ma:1995ey}
\begin{equation}
ds^{2}=a^{2}\left(\eta\right)\left[-\left(1+2\Phi\right)d\eta^{2}+\left(1-2\Phi\right)d\vec{x}^{2}\right]\,,
\end{equation}
where $\eta$ is the conformal time and $\Phi$ the gravitational
field. The energy momentum of a perfect fluid with energy density $\rho$, pressure $p$ and four-velocity
$u_{\mu}$ is given by
\begin{equation}
T_{\mu\nu}=\left(\rho+p\right)u_{\mu}u_{\nu}-pg_{\mu\nu}\,,\label{eq:energy-mom-tensor}
\end{equation}
which include background (overbar quantities) plus perturbed quantities:
$\rho=\bar{\rho}+\delta\rho$, $p=\bar{p}+\delta p$, $u_{\mu}=\bar{u}_{\mu}+v_{\mu}$.

We also restrict the analysis to the matter dominated era and the
actual DE dominated phase. Then, at background level, we have the Friedman
equations including matter (barions plus dark matter, indicated by the subscript $m$) and dark energy (indicated by the subscript $de$) given by
\begin{equation}
\mathcal{H}^{2}\equiv\left(\frac{a'}{a}\right)^{2}=\frac{8\pi G}{3}a^{2}\left(\bar{\rho}_{m}+\bar{\rho}_{de}\right)\,,\label{eq:friedman-1}
\end{equation}
\begin{equation}
\mathcal{H}'=-\frac{4\pi G}{3}\left(\bar{\rho}_{m}+\bar{\rho}_{de}\left(1+w\right)\right)\,,\label{eq:friedman-2}
\end{equation}
where $w=\bar{p}_{de}/\bar{\rho}_{de}$ is the EoS parameter for
DE and the prime indicates derivative with respect to the conformal time, $\eta$. Density parameters of matter and DE are defined by
\begin{equation}
\Omega_m = \frac{\bar{\rho}_m(a)}{\rho_c(a)}\,\,\, \mbox{and}\,\,\,
\Omega_{de} = \frac{\bar{\rho}_{de}(a)}{\rho_c(a)}\,,
\end{equation}
where $\rho_c=\bar{\rho}_{m}+\bar{\rho}_{de}$ is the critial density.

In this review, we present examples using the CPL parametrization
of the EoS \citet{Chevallier:2000qy,Linder:2002et}
\begin{equation}
w=w_{0}+w_{a}\left(1-a\right)\,,\label{eq:CPL-parametrization}
\end{equation}
where $w_0$ and $w_a$ are constants.
Since $\Lambda$ gives a good description for the background
evolution, we fix $w_{0}=-1$ and will vary $w_{a}$ to show the impact
of DE fluctuations in a scenario that is very similar to $\Lambda$CDM
at low redshift.

In Fourier space, the $\left(00\right)$ component of Einstein equations
is given by
\begin{equation}
k^{2}\Phi+3\mathcal{H}\left(\Phi'+\mathcal{H}\Phi\right)=4\pi Ga^{2}\left(\bar{\rho}_{m}\delta_{m}+\bar{\rho}_{de}\delta_{de}\right)\,,\label{eq:einstein-00}
\end{equation}
where $\delta_{\ell}=\delta\rho_{\ell}/\bar{\rho}_{\ell}$ is the
density contrast for a given component identified by the subscript
$\ell$. The conservation equations for perturbations of each fluid
component, $\nabla^{\nu}\delta T_{\nu}^{\mu}=0$, can be written as
\begin{equation}
\delta'+3\mathcal{H}\left(\delta p/\delta\rho-w\right)\delta+\left(1+w\right)\left(\theta-3\Phi'\right)=0\label{eq:continuity-newt-gauge}
\end{equation}
\begin{equation}
\theta'+\mathcal{H}\left(1-3c_{a}^{2}\right)\theta=k^{2}\Phi+\frac{\delta p/\delta\rho k^{2}\delta}{1+w}\,,\label{eq:euler-newt-gauge}
\end{equation}
where $\theta =ik^jv_j$ is the divergence of the fluid peculiar velocity and $c_{a}^{2}=\bar{p}'/\bar{\rho}'$ is the adiabatic sound speed
of the fluid. The pressure perturbation is given by \citet{Bean2004}
\begin{equation}
\delta p=c_{s}^{2}\delta\rho+3\mathcal{H}\left(1+w\right)\left(c_{s}^{2}-c_{a}^{2}\right)\bar{\rho}\frac{\theta}{k^{2}}\,,\label{eq:pressure-pert}
\end{equation}
where $c_{s}^{2}=\left(\delta p/\delta\rho\right)_{{\rm rest}}$ is
the sound speed of the fluid in its rest frame.

An adiabatic or barotropic fluid is defined by $c_{s}=c_{a}$. The
fluid is said to possess entropy perturbation if $c_{s}\neq c_{a}$.
In general, DE models have entropy perturbations, but it is also possible
to construct adiabatic models, e.g., \citet{Linder2009,Unnikrishnan2010}

Let us analyze the behaviour of perturbations in an adiabatic DE model,
which has simpler equations -- see \citet{Ballesteros2010b} for
a more general treatment. In this case, the last term in (\ref{eq:pressure-pert})
vanishes. Assuming $w$ and $c_{s}$ as constants, from (\ref{eq:continuity-newt-gauge})-(\ref{eq:pressure-pert})
we can determine a second order equation for DE density contrast
\begin{multline}
\delta_{de}''+\mathcal{H}\left(1-3w\right)\delta_{de}'+\left[3\mathcal{H}'\left(c_{s}^{2}-w\right)+3\mathcal{H}^{2}\left(c_{s}^{2}-w\right)\left(1-3c_{s}^{2}\right)+c_{s}^{2}k^{2}\right]\delta_{de}\\
=-\left(1+w\right)\left[k^{2}\Phi-3\mathcal{H}\left(1-3c_{s}^{2}\right)\Phi'-3\Phi''\right]\,.\label{eq:delta-de-second-order}
\end{multline}

Let us focus on small scales, $k^{2}\gg\mathcal{H}^{2},\mathcal{H}'$
and $k^{2}\Phi\gg\mathcal{H}\Phi',\Phi''$. During the matter dominated
era, we have the well-known solution $\delta_{m}\propto a$ and $\Phi=\text{constant}$,
and equation (\ref{eq:delta-de-second-order}) simplifies to
\begin{equation}
\delta_{de}''+\mathcal{H}\left(1-3w\right)\delta_{de}'+c_{s}^{2}k^{2}\delta_{de}=-\left(1+w\right)k^{2}\Phi\,.\label{eq:delta-de-second-order-small-scale}
\end{equation}
For non-negligible $c_{s}^{2}$, equation (\ref{eq:delta-de-second-order})
has a constant solution
\begin{equation}
\delta_{de}=-\frac{\left(1+w\right)}{c_{s}^{2}}\Phi\,.\label{eq:delta-de-cs}
\end{equation}
Since, from (\ref{eq:einstein-00}), $\delta_{m}\sim k^{2}\Phi$,
DE perturbations are usually negligible on small scales when compared
to the matter perturbations.

For negligible $c_{s}$, equation (\ref{eq:delta-de-second-order})
has the following solution \citet{Abramo2009,Sapone2009,Creminelli2009}
\begin{equation}
\delta_{de}=-\frac{\left(1+w\right)}{\left(1-3w\right)}\delta_{m}\,,\label{eq:delta-de-null-cs}
\end{equation}
which is a good approximation even for Early DE models \citet{Batista:2013oca}.

As can be seen, the magnitude of DE perturbations is determined by
both $w$ and $c_{s}$. If $w\simeq-1$, as should be at low-$z$,
DE perturbations are strongly suppressed regardless of $c_{s}$. For
DE models in which $w$ deviates from $-1$ at higher redshift and
with low or negligible $c_{s}$, DE perturbations can be as large
as matter perturbations in the matter-dominated era, as indicated by
(\ref{eq:delta-de-null-cs}). As we will see, in this case, DE fluctuations
can become nonlinear and impact structure formation.

The solution (\ref{eq:delta-de-null-cs}) also indicates a correlation
between matter and DE perturbations. Positive matter perturbations,
which will be associated with the formation of halos in the nonlinear
regime, will induce DE overdensities in the case of $1+w>0$ and underdensities
if $1+w<0$. As we will see later, in the nonlinear regime, this correlation
can induce a pathological behaviour for phantom DE models ($w<-1$),
namely, $\delta_{de}<-1$.

The comoving scale bellow which DE perturbations are strongly suppressed
by its pressure support is given by the sound horizon
\begin{equation}
r_{s}=\int_{a_{i}}^{1}\frac{c_{s}da}{a^{2}H}\,,\label{eq:sound-horizon}
\end{equation}
here $H=\dot{a}/a$ is the usual Hubble parameter and the dot indicates derivative
with respect to time. 
In regions below this scale, the pressure support halts the growth
of DE perturbations. As we will see, quintessence models have $c_{s}=1$,
then $r_{s}\sim1/H_{0}$ and their perturbations are very small on
scales below the Hubble radius.

Let us estimate the value of $c_{s}$ that induce large DE perturbations
on small scales. For instance, $c_{s}=10^{-3}$ gives a sound horizon
$r_{s}\simeq14\text{Mpc}$ (assuming best fit Planck18 cosmology \citet{Aghanim2020}).
This value is associated with a mass scale 
$M=4\pi/3\bar{\rho}_{m0}\left(r_{s}/2\right)^{3}\simeq2\times10^{14}M_{\odot}$, where $\rho_{m0}$ is the matter density now. The value $c_{s}=10^{-3}$ was indeed reported to impact the formation
of halos of such mass by \citet{Basse2011}. As we will discuss later,
in this work the authors considered that DE fluctuations do not break
linear approximation. In the limit of $c_{s}\rightarrow0$, however,
DE fluctuations can become nonlinear, depending on the value of $w$.
Nevertheless, we can consider that impact of DE fluctuations to become
important for nonlinear structure formation for $c_{s}<10^{-3}$.

\section{Dark Energy Models}

Now that we have understood under which circumstances DE perturbations
can be significant on small scales, let us discuss some DE 
models that can present such large perturbations. The variety of DE
models is enormous, with several reviews, books and extensive analysis
about them, e.g., \citet{Copeland:2006wr,luca_book,Yoo2012,Tsujikawa2013a,Ade2016}.
We will focus on the k-essence class of models, which provide enough generality
for $w$ and $c_{s}$ to permit large DE perturbations.

Models of k-essence were initially introduced in the context of inflation
\citet{ArmendarizPicon1999,Garriga:1999vw} and shortly after applied
to describe DE \citet{ArmendarizPicon:2000ah,Erickson2002}. A cosmological
model with k-essence is described by the following action 
\begin{equation}
S=\int d^{4}x\sqrt{-g}\left[-\frac{\mathcal{R}}{2\kappa^{2}}+\mathcal{L}\left(X,\varphi\right)\right]+S_{m}\,,
\end{equation}
where $\kappa^{2}=8\pi G$, $\mathcal{R}$ is the Ricci scalar, $X=-\frac{1}{2}g^{\mu\nu}\partial_{\mu}\varphi\partial_{\mu}\varphi$,
$\mathcal{S}_{m}$ the action for matter fields (photons, neutrinos,
baryons, dark matter) and $\mathcal{L}\left(X,\varphi\right)$ the
Lagrangian for the k-essence field, $\varphi$. The k-essence energy-momentum
tensor is given by
\begin{equation}
T_{\mu\nu}=\mathcal{L}_{,X}\nabla_{\mu}\varphi\nabla_{\nu}\varphi-\mathcal{L}g_{\mu\nu}\,,
\end{equation}
where subscript $,X$ represents the derivative with respect to $X$.
Comparing it to the perfect fluid energy momentum tensor (\ref{eq:energy-mom-tensor}),
we find that
\begin{equation}
\rho=2X\mathcal{L}_{,X}-\mathcal{L}\,,\,\,\,p=\mathcal{L}\,\text{ and }u_{\mu}=\frac{\nabla_{\mu}\varphi}{\sqrt{2X}}\,.
\label{eq:rho-p-v-k-ess}
\end{equation}
The sound speed of k-essence perturbations is given by \citet{Garriga:1999vw}
\begin{equation}
c_{s}^{2}=\frac{p_{,X}}{\rho_{,X}}\,.\label{eq:k-ess-sound-speed}
\end{equation}
In this framework it is clear that one has the freedom to choose the
functions $w$ and $c_{s}$. Next, let us have a look at some specific
popular realizations of DE models.

\subsection{Quintessence}

In the k-essence language, quintessence is defined by $\mathcal{L}=X-V\left(\varphi\right)$,
then its EoS parameter, defined by $p/\rho$ given by (\ref{eq:rho-p-v-k-ess}), and sound speed, given by (\ref{eq:k-ess-sound-speed}), read
\begin{equation}
w=\frac{X-V}{X+V}\,\text{ and }c_{s}^{2}=1\,.\label{eq:quintessence-EoS-cs}
\end{equation}
If the kinetic energy of the field dominates over its potential energy,
we have $w\simeq1$. In the opposite case, $w\simeq-1$. Its sound
speed, however, is always the same. Hence, although $w$ can be far
from $-1$, the sound speed of quintessence does not allow for large
perturbations on small scales.

This fact can also be understood via the Klein-Gordon equation for
quintessence perturbations, which can be written as \citet{Hwang:2001ua}
\begin{equation}
\delta\ddot{\varphi}+3H\delta\dot{\varphi}+\left(c_{s}^{2}k^{2}+\frac{d^{2}V}{d\varphi^{2}}\right)\delta\varphi=\dot{\varphi}\delta_{m}\,,\label{eq:KG-equation}
\end{equation}
where we explicitly introduced the the parameter $c_{s}$. Quintessence
models must have $V\sim H_{0}^{2}$ in order to accelerate the Universe
expansion recently, where $H_0=H(t_0)$ is the Hubble constant. Thus, for scales well bellow the Hubble radius,
$c_{s}^{2}k^{2}\gg\frac{d^{2}V}{d\varphi^{2}}$, and the field perturbations
are strongly suppressed on these scales.

Since quintessence was initially the most popular alternative model
to $\Lambda$, initially, many cosmologists considered DE perturbations
were effectively negligible on small scales. This picture started
to change with the appearance of new DE models which can have $c_{s}<1$.

\subsection{Tachyon}

A well-known model for DE that makes use of non-canonical scalar field
is the tachyon field. It was introduced in the context of string theory \citet{Sen:2002nu,Sen:2002an}, but can also be understood as a generalization of the relativistic particle lagrangian, \citet{Padmanabhan:2002sh}. The tachyon model is defined by the following lagrangian
\begin{equation}
\mathcal{L}=-V\left(\varphi\right)\sqrt{1-2X}\,,\label{eq:tachyon-lagrangian}
\end{equation}
Comological studies of this model include \citet{Padmanabhan:2002cp,Padmanabhan:2002sh,Bagla:2002yn,Abramo:2003cp}.
The EoS parameter and its sound speed are given by
\begin{equation}
w=2X-1\,\text{ and }c_{s}^{2}=-w\label{eq:tachyon-EoS-cs}
\end{equation}
The accelerated expansion occurs when $X\ll1$, thus $w\simeq-1$
and $c_{s}^{2}\simeq1$. In this case, the behaviour is very similar
to quintessence. However, earlier in the cosmological history, the
kinetic term can be larger, which yields a smaller $c_{s}$. In this
case, both $w$ and $c_{s}$ provide the conditions for large DE perturbations.

Considering a power-law potential, $V\left(\varphi\right)\propto\varphi^{-\alpha}$,
the actual impact of tachyon perturbations on CMB is small because
the accelerated expansion requires $\alpha\simeq0$, which, in turn,
suppresses the field perturbations \citet{Abramo:2004ji}. Nevertheless,
this model clearly shows that DE perturbations are not necessarily
small on scales below the Hubble radius.

For a constant potential, the tachyon model is equivalent
to the Chaplygin gas \citet{Kamenshchik:2001cp}, with equation of
state
\begin{equation}
p=-\frac{A}{\rho}\,,\label{eq:chap-EoS}
\end{equation}
where $A$ is a constant. This model and its generalized version was
studied by many authors, e.g., \citet{Bento:2002ps,Sandvik:2002jz,Makler2003,Bento2003,Amendola2003,Reis:2004hm}.
Both the tachyon and Chaplygin models gave rise to a scenario in which
DE and dark matter can be considered manifestations of a single component,
which became known as quartessence or unified dark energy \citet{Sahni2000a,Bilic:2001cg},
for a review of these ideas, see \citet{Bertacca2010}.

\subsection{Clustering DE}

It is possible to construct DE models with negligible sound speed.
In the context of k-essence, a model with a constant arbitrary $c_{s}$
is given by \citet{Kunz2015}
\begin{equation}
\mathcal{L}=M^{4}\left(\frac{X}{M^{4}}\right)^{\frac{1+c_{s}^{2}}{2c_{s}^{2}}}+V\left(\varphi\right)\,,
\end{equation}
where $M$ is a constant mass scale. In the context of quartessence,
\citet{Scherrer2004} proposed a model with very low $c_{s}$. A possible
issue when building such models is that distinct lagrangians can have
the same $w$ and $c_{s}$, \citet{Unnikrishnan2008}.

In effective field theory, models with $c_{s}^{2}\sim-10^{-30}$ were
described by \citet{Creminelli2009}. Although negative $c_{s}^{2}$
yields gradient instabilities, this value is so small that no relevant
effect on cosmological scales is expected. This work also concludes
that no pathological behaviour is present for phantom models, $w<-1$.
However, as we will show later on, in the nonlinear regime, it is
possible that phantom DE with negligible sound speed can present $\delta_{de}<-1$,
i.e., a pathological situation with negative energy density, $\rho_{de}=\bar{\rho}_{de}(1+\delta _{de}$).

Models with identically zero sound speed using two scalar fields were
proposed by \citet{Lim2010}. In Horndeski theories, there are even
more possible realizations of DE with low sound speed, which can be
chosen as a function of four physical parameters \citet{Bellini2014}.

Given the variety of possible CDE models, the phenomenological implementation
of parametrizations of $w$ and $c_{s}$ is valuable to explore the
possible impact of DE perturbations on structure formation. As already
mentioned, we will show examples using $w=w_{0}+w_{a}\left(1-a\right)$
and constant $c_{s}$.

\section{The spherical collapse model}
Once we know the basic behaviour of DE linear perturbations, we may ask how they impact the 
structure formation on small scales. The SCM is an analytical approach first proposed to study the nonlinear evolution of matter perturbations in the Universe. It can be used in connection to analytic and semi-analytic halo mass functions to determine the abundance of matter halos, which we will describe in section 7. In this section, we will review the classic formulation of the SCM and the main quantities of cosmological interest. We also discuss some generalizations capable of treating homogeneous DE and the first proposed model that considers the nonlinear evolution of DE fluctuations. Later on, we will present a more general framework for the SCM and a detailed discussion about threshold density, which is an important quantity that determines the abundance of halos. 

\subsection{Einstein-de-Sitter Universe}

The SCM was first proposed by \citet{Gunn:1972sv}. The important
conclusions for our porpouses are described in \citet{Padmanabhan}
and \citet{Sahni1995}. This model describes the dynamics of spherical
shell of radius $R$ that encloses a mass $M_{m}$. The dynamical
equation for the shell radius is 
\begin{equation}
\ddot{R}=-\frac{GM_{m}}{R^{2}}\, ,\label{eq:sc-classic}
\end{equation}
where the dot indicate derivative with respect to time. 
The mass is given by
\begin{equation}
M_{m}=\frac{4\pi}{3}R^{3}\bar{\rho}_{m}\left(1+\delta_{m}\right)\,,\label{eq:mass-matter}
\end{equation}
where $\delta_{m}$ is assumed to be the averaged density contrast
inside the sphere of radius $R$, which can also be understood as
a top-hat profile for the matter fluctuation. The total mass is supposed
to be conserved, $dM_{m}/dt=0$, which yields the following equation
for the matter contrast
\begin{equation}
\dot{\delta}_{m}+3\left(\frac{\dot{R}}{R}-\frac{\dot{a}}{a}\right)\left(1+\delta_{m}\right)=0\,,\label{eq:continuty-SC}
\end{equation}
which has the solution 
\begin{equation}
\delta_{m}+1=\left(\delta_{mi}+1\right)\left(\frac{a}{a_{i}}\frac{R_{i}}{R}\right)^{3}\,,
\label{eq:delta-m-SC-solution}
\end{equation}
where the subscripts $i$ indicate the initial values.

The first integral of (\ref{eq:sc-classic}) is given by
\begin{equation}
\frac{1}{2}\dot{R}^{2}-\frac{GM_{m}}{R}=E\,,\label{eq:first-integral-R}
\end{equation}
where $E$ is a constant of integration identified as the total energy
of the shell -- the kinetic energy ($K=\dot{R}^{2}/2$) plus the
potential energy per unit of mass ($U=-GM_{m}/R$). Note that, in
the derivation of (\ref{eq:first-integral-R}), the fact that $M_{m}$
is constant was used. As we will see later on, in the presence of
DE, the effective mass inside the shell is not conserved anymore.

Assuming that the shell initially expands with the background, $R\propto a$,
we have $\dot{R}_{i}=H_{i}R_{i}$. Using (\ref{eq:mass-matter}),
the total energy can be expressed as
\begin{equation}
E=\frac{\left(H_{i}R_{i}\right)^{2}}{2}\left[\frac{1}{\Omega_{mi}}-\left(1+\delta_{mi}\right)\right]\,,\label{eq:total-energy-SC}
\end{equation}
where $\Omega_{mi}=\Omega_{m}\left(a_{i}\right)$. For $1+\delta_{mi}>1/\Omega_{mi}$,
the total energy of the shell is negative. Thus it is expected that
the initially expanding dust cloud will grow to a maximum radius,
contract and eventually form a bounded structure. In EdS Universe,
$\Omega_{m}=1$, and any region with positive $\delta_{mi}$ will
eventually collapse. As we will see next, this is indeed the behaviour
given by solving for the time evolution of $R$, which will describe
the formation of a matter halo. On the other hand, if $1+\delta_{mi}<1/\Omega_{mi}$,
the total energy of the shell is positive and the cloud will expand
forever, forming a void with $-1<\delta_{m}<0$.

The maximum shell radius is given by $\dot{R}=0$ and is called the
turn-around radius, given by
\begin{equation}
R_{ta}=R_{i}\frac{1+\delta_{mi}}{1+\delta_{mi}-1/\Omega_{mi}}\,.
\label{eq:turn-around-radius}
\end{equation}
This is indeed a maximum radius because equation (\ref{eq:sc-classic})
indicates that $\ddot{R}<0$. This fact also implies that no solution
of a minimal radius exists in the usual SCM, and consequently,
the model can not describe an equilibrium configuration. This shortcoming
is usually circumvented by assuming that the system achieves virial
equilibrium at some point of its evolution and that this state represents
the final bounded structure. There were efforts in implementing virialization
in the dynamics of the SCM, see \citet{Engineer:1998um,Shaw:2007tr}.

Equation (\ref{eq:sc-classic}) can be solved analytically with the
following parametrization
\begin{equation}
R=A\left(1-\cos x\right)\text{ and }t=B\left(\theta-\sin x\right)\,,\label{eq:sc-solution}
\end{equation}
where $A=R_{ta}/2$, $B=t_{ta}/\pi$. Substituting these expressions
in (\ref{eq:sc-classic}) one gets the relation $A^{3}=GM_mB^{2}.$

As can be seen from (\ref{eq:sc-solution}), the shell radius is maximum
at $\theta=\pi$ (turn-around) and then begins to shrink. Formally,
as $x\rightarrow2\pi$, $R\rightarrow0$ and, from (\ref{eq:delta-m-SC-solution}), $\delta_{m}\rightarrow\infty$.
This is identified as the moment of collapse. In reality, these values
are not achieved because, for a given shell approaching collapse,
other inner shells containing collisionless matter have already
crossed each other. Thus, the total mass within a specific radius is
not conserved anymore, signalling the break down of the model. Nevertheless,
the collapse time is used to define quantities used to estimate
the abundance of halos.

Using conservation of mass, given by equation (\ref{eq:mass-matter}),
and the background density evolution in EdS $\bar{\rho}_{m}=\left(6\pi Gt^{2}\right)^{-1}$,
the nonlinear evolution of matter fluctuations is given by
\begin{equation}
\frac{\rho_{m}}{\bar{\rho}_{m}}=1+\delta_{m}^{NL}=\frac{9}{2}\frac{\left(x-\sin x\right)^{2}}{\left(1-\cos x\right)^{3}}\,,\label{eq:SC-nonlinear-delta}
\end{equation}
where the superscript $NL$ indicates the nonlinear value of the density constrast.
Expanding for small $x$, we get the linear evolution
\begin{equation}
\delta_{m}^{L}=\frac{3}{5}\left(\frac{3}{4}\right)^{2/3}\left(x-\sin x\right)^{2/3}\,,\label{eq:SC-linear-delta}
\end{equation}
where the superscript $L$ indicates the linear value of the density contrast.
At turn-around, we have 
\begin{equation}
\delta_{m}^{NL}\left(x=\pi\right)=\frac{9\pi^{2}}{16}-1\simeq4.552\label{eq:delta_NL-turn-around}
\end{equation}
and
\begin{equation}
\delta_{m}^{L}\left(x=\pi\right)=\frac{3}{5}\left(\frac{3\pi}{4}\right)^{2/3}\simeq1.062\,.\label{eq:delta_LIN-turn-around}
\end{equation}
As can be seen, at this time, the linear evolution gives a density
contrast of order one, indicating the transition from linear to nonlinear
regime. 

The collapse threshold (or critical density contrast) is the extrapolated
linear overdensity value above which a bound structure is considered
formed, a halo. Usually, it is defined at the collapse time
\begin{equation}
\delta_{c}\equiv\delta_{m}^{L}\left(x=2\pi\right)=\frac{3}{5}\left(\frac{3\pi}{2}\right)^{2/3}\simeq1.686\,.\label{eq:delta_c-classic}
\end{equation}
As we will see later, the value $\delta_{c}$ is of great importance
in structure formation studies that make use of analytic or semi-analytic
mass functions. This quantity is modified in the presence of homogeneous and inhomogeneous DE.

It is also important to describe the state of equilibrium of the halo,
which is assumed to obey the virial theorem for non-relativistic particles.
Virialization occurs when $U\left(R_{{\rm v}}\right)=-2K\left(R_{{\rm v}}\right)$,
were $R_{{\rm v}}$ is the virialization radius. At turn-around, $E=U\left(R_{ta}\right)=U\left(R_{{\rm v}}\right)+K\left(R_{{\rm v}}\right)$,
then we get the relation
\begin{equation}
R_{{\rm v}}=\frac{R_{m}}{2}\,.\label{eq:r-vir}
\end{equation}
From this, we can compute the virialization time ($x=3\pi/2$)
\begin{equation}
t_{{\rm v}}=\left(\frac{3}{2}+\frac{1}{\pi}\right)t_{ta}\simeq1.818t_{ta}\,.\label{eq:t-vir}
\end{equation}
The collapse time ($x=2\pi$) is 
\begin{equation}
t_{coll}=2t_{m}\,.\label{eq:t-coll}
\end{equation}
With these quantities, two definitions of virialization overdensity
are usually found in the literature. One possibility is define the
virialization overdensity with respect to the matter background density
\begin{equation}
\Delta_{{\rm cm}}\equiv\frac{\rho_{m}\left(t_{v}\right)}{\bar{\rho}_{m}\left(t_{c}\right)}=18\pi^{2}\simeq177.7\,.\label{eq:Dvir-coll-matter}
\end{equation}
Another common choice uses the critical density as reference
\begin{equation}
\Delta_{{\rm cc}}\equiv\frac{\rho_{m}\left(t_{v}\right)}{\bar{\rho}_{c}\left(t_{c}\right)}\,.\label{eq:Dvir-coll-crit}
\end{equation}
Of course, $\Delta_{{\rm cm}}$ and $\Delta_{{\rm cc}}$ give essentially
the same results in a matter dominated Universe, but they differ when
DE becomes important, see figure \ref{fig:D-virialization}.

The overdensity $\Delta\thickapprox178$ became a reference value
used in N-body simulations in order to identify halos, but many other
definitions can be used, see \citet{Despali2016} for an analysis
of several such choices. Interestingly, this paper has shown that
the use of the quantity $\Delta_{{\rm cc}}$ yields more universal
mass functions, in the sense of its dependence on redshift and cosmological
parameters. We will return to this discussion when dealing with the
impact of CDE on halo mass functions.

Yet another possibility is to compute both the total and background
densities at the virialization time, as suggested by \citet{Lee2010b}.
In this case the virialization overdensity is given by
\begin{equation}
\Delta_{{\rm v}}\equiv\frac{\rho_{m}\left(t_{v}\right)}{\bar{\rho}_{m}\left(t_{v}\right)}=\frac{9}{2}\left(\frac{3\pi}{2}+1\right)^{2}\simeq146.8\,.\label{eq:Dvir-vir}
\end{equation}
The value of $\Delta_{{\rm v}}$ is slightly smaller than $\Delta_{{\rm cc}}$.
Later on, will show that in the presence of DE these two quantities
decay with redshift, whereas $\Delta_{{\rm cm}}$ grows (see figure \ref{fig:D-virialization}). Analogously,
we can define the threshold density at virialization
\begin{equation}
\delta_{{\rm v}}\equiv\delta_{m}^{L}\left(x=\frac{3\pi}{2}\right)=\frac{3}{5}\left(\frac{3}{4}\right)^{2/3}\left(\frac{3\pi}{2}+1\right)^{2/3}\simeq1.583\,.\label{eq:critical-delta-virial}
\end{equation}

\subsection{Spherical collapse model with homogeneous dark energy\label{subsec:SC-model-homo}}

In the presence of homogeneous DE, the SCM has to be modified
to include the effects of this new component on the dynamics of the
shell radius. Now we have
\begin{equation}
\ddot{R}=-\frac{G}{R^{2}}\left(M_{m}+M_{de}\right)\,,\label{eq:sc-classic-hom-de}
\end{equation}
where
\begin{equation}
M_{de}=\frac{4\pi}{3}R^{3}\bar{\rho}_{de}\left(1+3w\right)\label{eq:de-hom-mass}
\end{equation}
is the effective mass associated with homogeneous DE. This definition
can be understood via the Poisson equation in the presence of a relativistic
fluid
\begin{equation}
\nabla^{2}\Phi=4\pi G\left(\rho+3p\right)\,.\label{eq:poisson}
\end{equation}

The first study of the equation (\ref{eq:sc-classic-hom-de}) was
done by \citet{Lahav:1991wc} considering $\Lambda$ as a DE component.
It was found that the ratio of virial to turn-around radius is smaller
in the presence of $\Lambda$, given by the following expression
\begin{equation}
\frac{R_{{\rm v}}}{R_{{\rm ta}}}\simeq\frac{1-\eta/2}{2-\eta/2}\,,\label{eq:r-vir-Lahav}
\end{equation}
where $\eta=\Lambda/4\pi G\bar{\rho}_{m}\left(t_{ta}\right)$. Thus,
in the presence of $\Lambda$, the halo has to contract more with
respect to EdS to achieve the virial equilibrium. This indicates that
halos formed in the late Universe have a distinct structure than those
formed at high-$z$, when DE was very subdominant.

Other authors have studied the SCM in the presence of $\Lambda$,
e.g., \citet{Lacey1993,Kitayama1996}. Although analytical solutions
for $\delta_{c}$ and $\Delta_{cc}$ were found, the following fits
presented by \citet{Kitayama1996} became popular:
\begin{equation}
\delta_{c}\simeq\frac{3\left(12\pi\right)^{2/3}}{20}\left(1+0.0123\log_{10}\Omega_{m}\left(z\right)\right)\label{eq:delta-c-KS}
\end{equation}
and
\begin{equation}
\Delta_{cm}\simeq18\pi^{2}\left(1+0.4093w_{f}^{0.9052}\left(z\right)\right)\,,\label{eq:delta-v-KS}
\end{equation}
where $w_{f}\left(z\right)=1/\Omega_{m}\left(z\right)-1$. Expression
(\ref{eq:delta-c-KS}) shows that the the influence of $\Lambda$
on the critical threshold is very small, giving $\delta_{c}\left(z=0\right)\simeq1.676$
for $\Omega_{m0}=0.3$ (only $0.64\%$ different than the EdS value).
However, the change in virialization overdensity is much larger. We
have $\Delta_{cm}\left(z=0\right)\simeq334.2$ for $\Omega_{m0}=0.3$,
about two times EdS value.

Studies of DE with constant $w$ were done by \citet{Wang:1998gt,Weinberg2003}.
The fitting functions proposed in the latter, shows variations of
$\delta_{c}\left(z\right)$ bellow $0.56\%$ for $-1<w<-0.8$ with
respect to the $\Lambda$CDM fit. For $\Delta_{cm}$ the differences
are lower than $17\%$. The differences can be slightly larger in
some quintessence models \citet{Mainini2003}

\begin{figure}
\centering{}\includegraphics[scale=0.6]{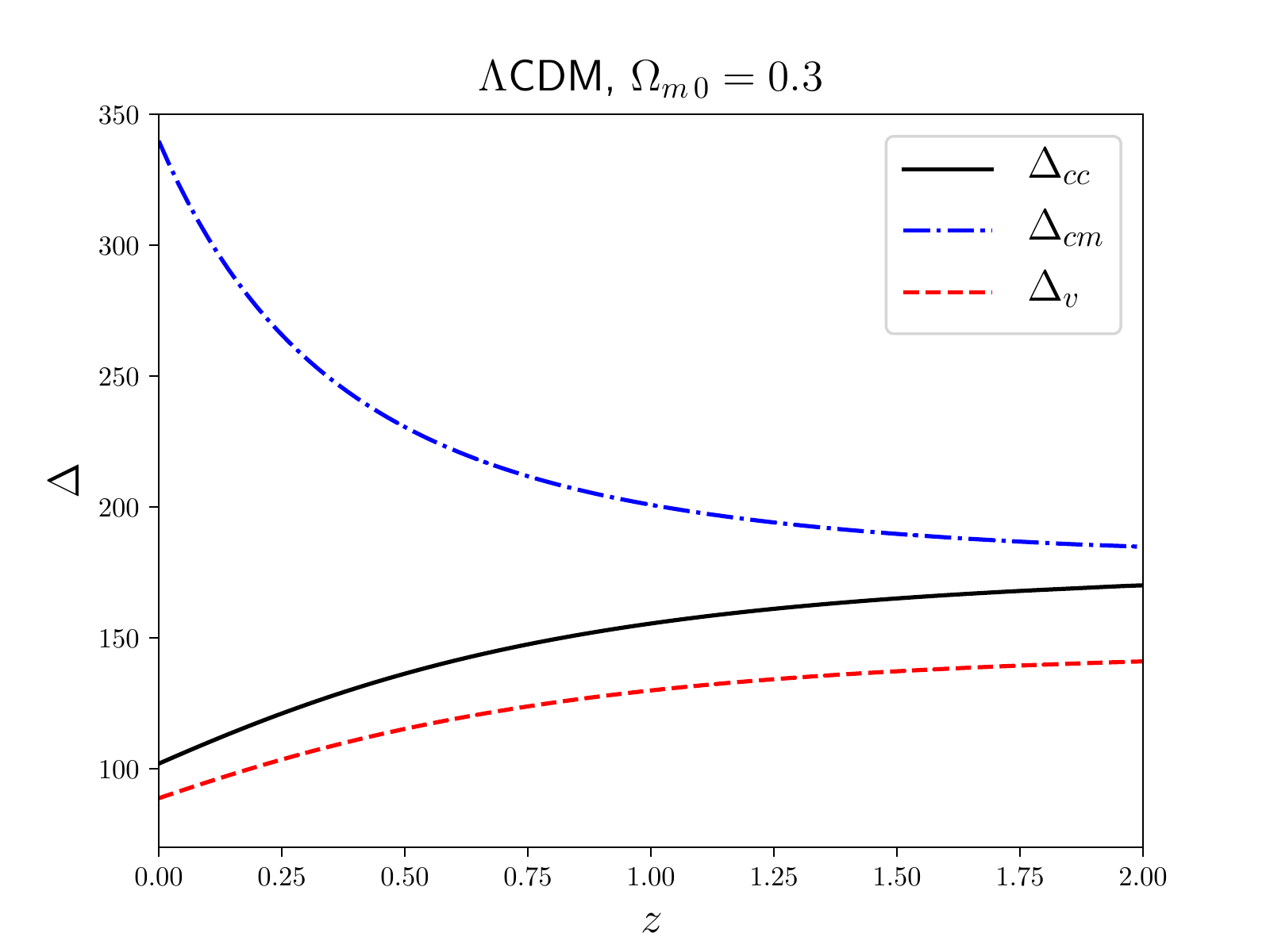}
\caption{Redshift evolution of three virial overdensities definitions, equations
(\ref{eq:Dvir-coll-matter}), (\ref{eq:Dvir-coll-crit}) and (\ref{eq:Dvir-vir})
for the $\Lambda$CDM model with $\Omega_{m0}=0.3$. In this plot,
$\Delta_{{\rm v}}$ is determined using the method summarized by equation
(\ref{eq:basse-virialization}), while $\Delta_{cm}$ and $\Delta_{cc}$
by the proper numerical method discussed in section (\ref{subsec:Collapse-threshold,})\label{fig:D-virialization}}
\end{figure}

In figure \ref{fig:D-virialization}, we show a plot of these three
different definitions of virial overdensity. As discussed, the variations
in $\delta_{c}$ due to homogeneous DE are very small and will be
shown together with the case of CDE in figure \ref{fig:delta_c}.

\subsection{Spherical collapse model with inhomogeneous dark energy}

The first study of SCM in the presence of CDE was done by \citet{Mota:2004pa}.
DE was modeled as quintessence field, which equation of motion in
background is given by
\begin{equation}
\ddot{\varphi}+3H\dot{\varphi}+\frac{dV}{d\varphi}=0\,.\label{eq:Mota-vB-field-bck}
\end{equation}
Remind that, in the context of quintessence models, the fluctuations
of such field must be tiny on small scales because $c_{s}=1$. The
authors have considered, however, that the field can cluster on small
scales, assuming that, inside the collapsing region, the it obeys
the following equation
\begin{equation}
\ddot{\varphi}_{c}+3\frac{\dot{R}}{R}\dot{\varphi}_{c}+\frac{dV}{d\varphi_{c}}=\frac{\Gamma}{\dot{\varphi}_{c}}\,,\label{eq:SC-quintessence}
\end{equation}
where
\begin{equation}
\Gamma=3\alpha\dot{\varphi}_{c}^{2}\left(\frac{\dot{R}}{R}-H\right)
\end{equation}
is a quantity that describes the flux of DE in the collapsing region.
If $\alpha=1$ DE evolves in the same way inside and outside the collapsing
region, then DE is homogeneous, and $\varphi_{c}=\varphi$. On the other hand, if $\alpha=0$
the field can evolve differently than in the background, allowing
for DE fluctuations. The impact on $\delta_{c}$ was computed in \citet{Nunes:2004wn},
where differences of a few per cent with respect to $\Lambda$CDM
and redshift dependent features associated with the evolution of $w$
were found.

Although this model was a breakthrough regarding the possibility of
nonlinear fluctuations of DE, the implementation of equation (\ref{eq:SC-quintessence})
is ad doc and formally inconsistent with the Klein-Gordon equation,
(\ref{eq:KG-equation}). Equation (\ref{eq:SC-quintessence}) can
be interpreted as (\ref{eq:KG-equation}) with $c_{s}=0$, where the
the gravitational coupling is encoded in the dynamics of $R$. Later
it was shown that the canonical scalar field indeed has negligible
perturbations on small scales \citet{Mota:2007zn,Wang:2009hg}. However,
its perturbations can be more important on voids, which are much larger
than halos.

Despite this inconsistency, this model presented some of the important
impacts of DE fluctuations on the nonlinear regime, which were later
confirmed by other studies. In particular, the following findings
can be highlighted:
\begin{enumerate}
\item The amount DE fluctuations strongly depends on the evolution of $w$.
\item DE fluctuations impact the nonlinear evolution of $\delta_{m}$ and
virialization of halos.
\item The local EoS of DE can be distinct from $w$ due to DE fluctuations.
\end{enumerate}

\subsection{Other generalizations of the spherical collapse model}

Many other generalizations of the SCM were developed in the literature,
including modifications due to coupled DE \citet{Manera:2005ct,Wintergerst2010a},
modified gravity \citet{Martino:2008ae,Schaefer2008,Schmidt2010,Brax:2010tj,Borisov2012,Barreira2013,Kopp2013,Lopes2018,Lopes2019,Frusciante2020},
varying vacuum models \citet{Basilakos2010}, shear and rotation \citet{DelPopolo:2012dq,Pace2014a,Mehrabi2017},
bulk viscosity \citet{Velten2014a}. In the next section we will present
a framework that can describe the nonlinear evolution in the CDE model.

\section{Spherical collapse model in the Pseudo-Newtonian Cosmology}

General Relativity is the standard theory to study gravitational phenomena
of fluids with relativistic pressure, $p\sim c^{2}\rho$. In the context
of structure formation, the gravitational fields are small, and the
weak field limit can be used. Therefore, one can consider using the
Newtonian theory, but taking into account the effects of relativistic
pressure. This approach was indeed used in \citet{MacCrea51,Harrison65}
to study the background evolution of the Universe. It was shown
that Newtonian equations with the correction for relativistic pressure
reproduce the Friedman equations.

However, the application of these modified Newtonian equations for
the study of cosmological perturbations is more controversial. \citet{Sachs:1967er}
showed that perturbations in fluids with $p\neq0$ do not agree with
the relativistic treatment. For about three decades, the use of Newtonian
equation for cosmology was halted until the origin of the discrepancy
reported in 1967 by Sachs and Wolfe was found and treated in \citet{Lima:1996at}.

According to \citet{Lima:1996at}, the system of equations (the subscript
$\ell$ identifies the different fluids under consideration), which
we call Pseudo-Newtonian Cosmology (PNC),
\begin{equation}
\frac{\partial\rho_{\ell}}{\partial t}+\vec{\nabla}\cdot\left(\rho_{\ell}\vec{u}_{\ell}\right)+\frac{p_{\ell}}{c^{2}}\vec{\nabla}\cdot\left(\vec{u}_{\ell}\right)=0\label{eq:PNC-1}
\end{equation}
\begin{equation}
\frac{\partial\vec{u}_{\ell}}{\partial t}+\left(\vec{u}_{\ell}\cdot\vec{\nabla}\right)\vec{u}_{\ell}=-\vec{\nabla}\Phi-\frac{\vec{\nabla}p_{\ell}}{\rho_{\ell}+\frac{p_{\ell}}{c^{2}}}\label{eq:PNC-2}
\end{equation}
\begin{equation}
\nabla^{2}\Phi=4\pi G\sum_{\ell}\left(\rho_{\ell}+\frac{3p_{\ell}}{c^{2}}\right)\label{eq:PNC-3}
\end{equation}
provides the correct growing modes for linear perturbations of fluids
with relativistic pressure. The PNC was extensively studied and applied
in the context of cosmological perturbations \citet{Reis:2003fs,Reis:2004hm,Hwang:2005xt,Fabris:2008hy,Abramo2009,Velten2013,Hwang2016}.
In particular, it was shown that PNC growing modes for DE perturbations
agrees with the relativistic analysis \citet{Abramo2009}.

The PNC can also be used to generalize the SCM \citet{Abramo2007,Abramo2009}.
The perturbed equations in comoving coordinates with background are
given by
\begin{equation}
\dot{\delta}_{\ell}+3H\left(c_{s\,\ell}^{2}-w_{\ell}\right)\delta_{\ell}+\left[1+w_{\ell}+\left(1+c_{s\,\ell}^{2}\right)\delta_{\ell}\right]\frac{\vec{\nabla}\cdot\vec{v}_{\ell}}{a}+\frac{\vec{v}_{\ell}\cdot\vec{\nabla}\delta_{\ell}}{a}=0\,,\label{eq:continuity-PNC}
\end{equation}
\begin{equation}
\dot{\vec{v}}_{\ell}+H\vec{v}_{\ell}+\frac{\vec{v}_{\ell}\cdot\vec{\nabla}}{a}\vec{v}_{\ell}=-\frac{\vec{\nabla}\Phi}{a}-\frac{c_{s\,\ell}^{2}\vec{\nabla}\delta_{\ell}}{a\left[1+w_{\ell}+\left(1+c_{s\,\ell}^{2}\right)\delta_{\ell}\right]}\,,\label{eq:euler-PNC}
\end{equation}
\begin{equation}
\frac{\nabla^{2}\Phi}{a^{2}}=4\pi G\sum_{\ell}\bar{\rho}_{\ell}\delta_{\ell}\left(1+3c_{s\,\ell}^{2}\right)\,,\label{eq:poisson-PNC}
\end{equation}
where $\vec{v}$ is the peculiar velocity of the fluid. Clearly, these
equations are more general then those of the usual SCM. Let us consider
the simplifying assumptions that will yield the correspondence between
them.

Remind that the SCM assumes a top-hat profile, thus we need $\vec{\nabla}\delta=0$
inside the shell. The other quantities must be consistent with this
assumption. Equation (\ref{eq:continuity-PNC}) must depend only on
time, thus $\vec{\nabla}\cdot\vec{v}=\theta\left(t\right)$. Taking
the divergence of (\ref{eq:euler-PNC}) we get
\begin{equation}
\dot{\theta}_{\ell}+H\theta_{\ell}+\frac{\theta_{\ell}^{2}}{3a}=-4\pi Ga\sum_{\ell}\bar{\rho}_{\ell}\delta_{\ell}\left(1+3c_{s\,\ell}^{2}\right)\label{eq:euler-theta-CPN}
\end{equation}

Now let us turn our attention to the parameter $c_{s}$. At first
glance, assuming that $\vec{\nabla}\delta=0$ and $\vec{\nabla}\cdot\vec{v}=\theta\left(t\right)$
seems compatible to any choice $c_{s}=c_{s}\left(t\right)$. However,
considering two distinct fluids, say pressureless matter ($w_{m}=c_{s\,m}=0$)
and DE ($w_{de}=w_{de}\left(t\right)$ and $c_{s\,de}=c_{s\,de}\left(t\right)$),
implies that each fluid has its own dynamical equation, (\ref{eq:euler-theta-CPN}).
This indicates that there is no unique spherical shell radius, because
the two fluids can flow with distinct velocities.

Another problem about using a generic function $c_{s}\left(t\right)$
arises when extending the analysis to regions outside the shell. For
a top-hat profile, $\delta$ is discontinuous at the edge of
the shell, then $\vec{\nabla}\delta$ is ill-defined at this
point. A more realistic realization is to assume a smooth decay of
$\delta$ around the edge. In this case, its gradient is well-defined
and there exists a non-null pressure gradient, $c_{s}^{2}\vec{\nabla}\delta$,
around the shell radius. This, again, would make the two fluids flow
with distinct velocities and, more drastically, disrupt the original
homogeneous top-hat-like profile.

Therefore, regarding the value of $c_{s}$, the equivalence of equations
(\ref{eq:continuity-PNC})-(\ref{eq:poisson-PNC}) with the usual
SCM is achieved only for $c_{s}=0$. In this case, all fluids
share the same dynamical equation and the evolution of the system
is described by
\begin{equation}
\dot{\delta}_{\ell}-3Hw_{\ell}\delta_{\ell}+\left(1+w_{\ell}+\delta_{\ell}\right)\frac{\theta}{a}=0\label{eq:cont-SC-CPN}
\end{equation}
\begin{equation}
\dot{\theta}+H\theta+\frac{\theta^{2}}{3a}=-4\pi Ga\sum_{\ell}\bar{\rho}_{\ell}\delta_{\ell}\label{eq:euler-SC-CPN}
\end{equation}
For pressureless matter, equation (\ref{eq:continuity-PNC}) yields
\begin{equation}
\dot{\delta}_{m}+\left(1+\delta_{m}\right)\frac{\theta}{a}=0\,,
\end{equation}
Comparing it with equation (\ref{eq:continuty-SC}), we identify that
the divergence of the peculiar velocity is given by
\begin{equation}
\frac{\theta}{a}=3\left(\frac{\dot{R}}{R}-\frac{\dot{a}}{a}\right)\,.
\end{equation}
Inserting this relation in (\ref{eq:euler-SC-CPN}), one obtains 
\begin{equation}
\frac{\ddot{R}}{R}=-\frac{4\pi G}{3}\sum_{\ell}\bar{\rho}_{\ell}\delta_{\ell}+\frac{\ddot{a}}{a}\,.
\end{equation}
For the EdS model, we recover the usual spherical collapse equation, (\ref{eq:sc-classic}),
(here as a function of the density contrast)
\begin{equation}
\frac{\ddot{R}}{R}=-\frac{4\pi G}{3}\bar{\rho}_{m}\left(1+\delta_{m}\right)\,.
\end{equation}
In the presence of CDE, we get
\begin{equation}
\frac{\ddot{R}}{R}=-\frac{4\pi G}{3}\left[\bar{\rho}_{m}\left(1+\delta_{m}\right)+\bar{\rho}_{de}\left(1+3w+\delta_{de}\right)\right]\,.\label{eq:dd-radius-matter-de}
\end{equation}
Particularizing to homogeneous DE models, $\delta_{de}=0$, we recover
the SC equation in \citet{Wang:1998gt}.

The system of equations (\ref{eq:cont-SC-CPN}) and (\ref{eq:euler-SC-CPN})
was also derived in \citet{Creminelli2010} using Fermi coordinates
in a relativistic framework. The consistency of equation
(\ref{eq:euler-SC-CPN}) with k-essence equation of motion was also
verified in this work.

Summarizing, the SCM can be generalized to include other fluids
with zero pressure perturbation, but allowing for non-zero background
pressure. This is just the case of CDE with EoS parameter $w$. The
equations governing the nonlinear evolution of the fluids are then
given by: 
\begin{equation}
\dot{\delta}_{m}+\left(1+\delta_{m}\right)\frac{\theta}{a}=0\,
\label{eq:cont-SC-CPN-mat}
\end{equation}
\begin{equation}
\dot{\delta}_{de}-3Hw\delta_{de}+\left(1+w+\delta_{de}\right)\frac{\theta}{a}=0\,,
\label{eq:cont-SC-CPN-de}
\end{equation}
\begin{equation}
\dot{\theta}+H\theta+\frac{\theta^{2}}{3a}=-4\pi Ga\left(\bar{\rho}_{m}\delta_{m}+\bar{\rho}_{de}\delta_{de}\right)\,.
\label{eq:euler-SC-CPN-mat-de}
\end{equation}

Note that, generic values of $c_s$ can be considered in the system of equations (\ref{eq:continuity-PNC})-(\ref{eq:poisson-PNC}). However, the differential equations become partial in this case, and the 
correspondence with the orginal SCM is lost. We stress that the system (\ref{eq:cont-SC-CPN-mat})-(\ref{eq:euler-SC-CPN-mat-de}) is valid only in the limit $c_s\rightarrow 0$. 

A slightly different approach to treat DE perturbations in the SCM was proposed in \citet{Basse2011}. In this work, DE perturbations
were described in the linear regime, but allowing for non-null sound
speed. Then, as discussed previously, DE perturbations can not maintain
the top-hat profile. The new idea in \citet{Basse2011} was to consider
the spherical region with a DE perturbation given by
\begin{equation}
\delta^{th}{}_{de}\left(t\right)=\frac{1}{2\pi^{2}}\int dkk^{2}W\left(kR\right)\delta_{de}^{L}\left(k,t\right)\,,\label{eq:delta-de-top-hat}
\end{equation}
where $W\left(kR\right)$ is the top-hat window function in Fourier
space, given by (\ref{eq:top-hat-window}), and $\delta_{de}^{L}\left(k,t\right)$
obeys the linearized equations (\ref{eq:continuity-PNC}) and (\ref{eq:euler-PNC})
in Fourier space. This approach has the advantage to allow the use
of arbitrary sound speed up to values that do not break the linear
approximation for DE perturbations. However, it can not be used in
the full clustering regime, $c_{s}\rightarrow0$, when DE fluctuations
can become nonlinear. Nevertheless, the approach of \citet{Basse2011}
is essential to understand the impact of $c_{s}$ in the nonlinear
evolution of matter fluctuations, showing that $\delta_{c}$ and $\Delta_{{\rm v}}$
become mass-dependent, which can be an observational signature of
DE fluctuations in the abundance of galaxy clusters.

\section{Density threshold definitions}

Now that we have described the governing equations for the nonlinear
evolution of matter and DE fluctuations, let us turn our attention
to the determination of the threshold density that will be used in
the Halo Mass Functions (HMF) in section (\ref{sec:Halo-mass-functions}). Although this
determination is analytic in the usual in the $\Lambda$CDM model,
in the presence of DE with general $w$ and its possible fluctuations,
a numerical computation is necessary, which may introduce some issues.

We will first discuss the calculation of the usual collapse threshold,
$\delta_{c}$, and then the alternative virialization threshold, $\delta_{{\rm v}}$.
While the former is historically the most used one, the latter seems
to be more consistent with the actual contribution of DE fluctuations
in the collapsing region and in better accordance with results from simulations.

\subsection{Collapse threshold, $\delta_{c}$\label{subsec:Collapse-threshold,}}

In order to determine $\delta_{c}$ for CDE models,
one has to solve numerically the system of equations (\ref{eq:cont-SC-CPN-mat})-(\ref{eq:euler-SC-CPN-mat-de})
and its linearized version. The initial conditions for matter assume
the EdS linear solution, $\delta_{m}\propto a$ and the corresponding
value for $\theta$; for CDE the solution (\ref{eq:delta-de-null-cs})
can be used.

Having solved the system, one has to determine a criterion to find
the moment of collapse. In the standard SCM, the collapse threshold
can be found analytically computing the value of the linearly evolved
density contrast, (\ref{eq:SC-linear-delta}) at the time of collapse,
$R\rightarrow0$ or $\delta^{NL}_m\rightarrow\infty$. This suggests
that, in general, the determination of $\delta_{c}$ can be done by defining
a numerical threshold value for $\delta^{NL}_m$, above which the halo
is considered to be formed.

However, this implementation can introduce a small error in the determination
of $\delta_{c}$. From equations (\ref{eq:SC-nonlinear-delta}) and
(\ref{eq:SC-linear-delta}), we can see that, for small $x$, the
linear and nonlinear values of matter contrast differ by $\mathcal{O}\left(x^{4}\right)$.
Thus, their initial values are slightly different. In the analytical
solution for $\delta_{c}$, Eq. (\ref{eq:delta_c-classic}), this
difference is naturally taken into account. If one neglects this difference
in the initial conditions, the resulting $\delta_{c}\left(z\right)$
presents a small spurious increase with redshift, moving away from
the EdS value at high-$z$. This problem was noted in \citet{Herrera2017}
and further discussed in \citet{Pace2017}.

The approach presented in \citet{Pace2017} is twofold: initiate the
numerical integration at very high-$z$ ($z\sim10^{5}$) and make
a change of variable, $\delta_m\rightarrow1/f$. The first measure diminishes
the difference between $\delta^{NL}_m$ and $\delta^{L}_m$ at the beginning
and, consequently, the spurious increase of $\delta_{c}$. The change
of variable minimizes the numerical error in the determination of
the moment of collapse.

Although this implementation give good results, the proper approach
to accurately determine $\delta_{c}$ numerically is to use the analytical
solutions (\ref{eq:SC-nonlinear-delta}) and (\ref{eq:SC-linear-delta})
to determine the linear and nonlinear initial conditions \citet{Batista2017}.
These analytical solutions assume that the peculiar velocity is zero
initially, but more general expressions can be found in \citet{Padmanabhan}.
This method, however, might not be appliable to cosmologies in which
$\Omega_{m}$ is not very close to $1$ at the beginning of the integration
of the equations, like in Early DE models.

\begin{figure}
\centering{}\includegraphics[scale=0.6]{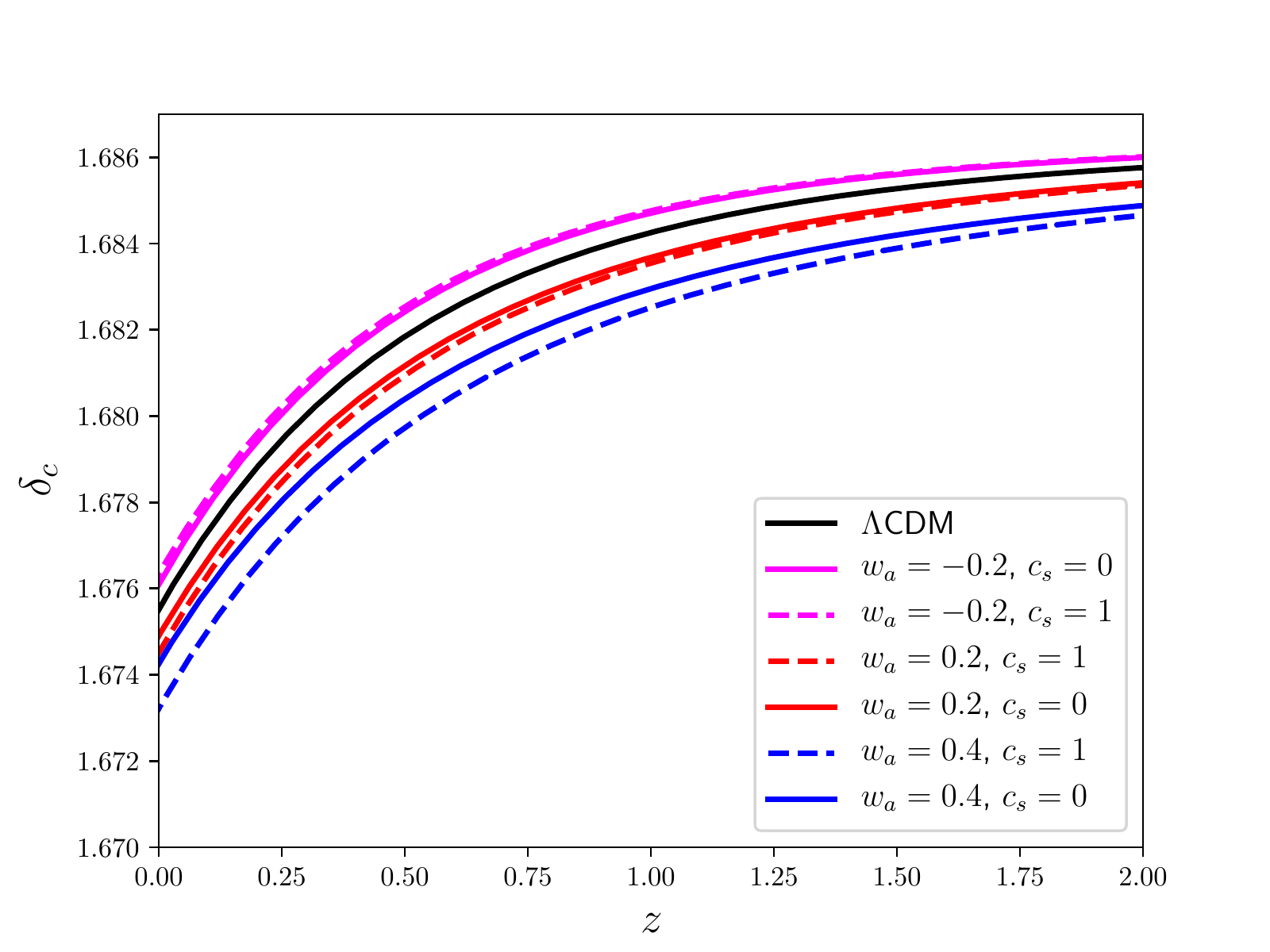}\caption{Collapse threshold as a function of redshift for various combinations
of $w_{a}$ and $c_{s}$ indicted in the plot legend. For all models,
we set $w_{0}=-1$ and $\Omega_{m0}=0.3$. As can be seen, the impact of $w_{a}$ and CDE
in this quantity is very small, below $1\%$ for the examples shown
in this plot.\label{fig:delta_c}}
\end{figure}

As we saw, the impact of homogeneous and CDE on $\delta_{c}$ is small, see also figure \ref{fig:delta_c}.
This suggests that the impact of DE fluctuations on structure formation
occurs mainly via modifications on the matter growth function. This
is somewhat unexpected because, as we will show later, at virialization,
DE fluctuations can account for up to $10\%$ of the total halo mass,
whereas the change in $\delta_{c}$ is below $1\%$. For possible contributions
of DE fluctuations to the halo mass, see \citet{Creminelli2010,Basse2012,Batista:2013oca}.

This insensitivity of the collapse threshold on DE fluctuations can
be understood as follows. First, $\delta_{c}$ is defined at the collapse
time, which numerically is implemented as a high value for $\delta_{m}^{NL}$.
Although $\delta_{de}^{NL}$ also grows, given the nature of the nonlinear
evolution, the higher value of $\delta_{m}^{NL}$ will grow exponentially
faster, thus making the DE contribution much less important at the
collapse time. Second, the collapse threshold is given solely by the
linear matter perturbation, which, although is impacted by DE perturbations,
does not consider the direct contribution of DE linear perturbations.

\subsection{Virialization threshold, $\delta_{{\rm v}}$}

It is important to note that, implicitly in the determination of $\delta_{c}$
just described lies the assumption that DE fluctuations are not directly
included in the quantities that define the collapse time and density
threshold, namely $\delta^{NL}_m$ and $\delta^{L}_m$. In the $\Lambda$CDM model, baryon and dark matter fluctuations
are the only relevant ones, besides a small contribution of massive
neutrinos \footnote{For the impact about massive neutrinos in the SCM, see \citet{Ichiki:2011ue,LoVerde2014}. Note that, since massive neutrinos can impact both the growth function and the collapse threshold, their effect can be degenerate with that from CDE models.}. Thus, the gravitational potential, which, for instance,
will deflect light rays of background galaxies or set up the potential
well that traps the hot intracluster gas, is entirely determined by
the fluctuations in dark matter and baryon components. In the presence
of DE fluctuations, the gravitational potential also depends on this
new type of inhomogeneity. Therefore, it would be natural to redefine
the threshold density, virial overdensity and growth function to take 
the contribution of DE fluctuations into account properly. For instance,
equations (\ref{eq:dd-radius-matter-de}) already suggests that the
effective mass inside a shell of radius $R$ includes DE fluctuations.

Then let us define effective quantities that include DE fluctuations.
The total mass inside a shell of radius $R$ is given by
\begin{equation}
M_{{\rm tot}}=M_{m}+M_{de}\,,\label{eq:total-matter}
\end{equation}
where $M_{de}=\frac{4\pi}{3}R^{3}\bar{\rho}_{de}\delta_{de}$. Usually
the background density of DE is not included in this mass definition,
e.g, \citet{Creminelli2009,Basse2012,Batista2017}. In the case of
CDE, its local EoS parameter
\begin{equation}
w_{c}=\frac{p_{de}}{\rho_{de}}=\frac{w}{1+\delta_{de}}
\end{equation}
gets less negative during the collapse. Consequently, locally,
DE becomes more similar to pressureless matter, \citet{Mota:2004pa,Abramo2008}. 
Such variations of the EoS may also be associated with soft-matter properties \citet{Saridakis:2021qxb}.

The fraction of DE mass in the halo at the virialization time is
\begin{equation}
\epsilon=\frac{M_{de}}{M_{m}}\,.
\end{equation}
Some authors have computed the values of $\epsilon$ with slightly
different approaches to determine the virialization time, \citet{Creminelli2010,Basse2011,Batista:2013oca,Batista2017,Heneka2018}.
In particular, using the approach summarized in equation (\ref{eq:basse-virialization}), which assumes
that CDE fluctuations behave as nonrelativistic particles, \citet{Batista2017} showed that $|\epsilon|$ can be up to $0.1$, depending on $w$. In the case
of phantom DE, this contribution is negative and positive for non-phantom.
These values of $\epsilon$ raise an interesting question: if CDE
can contribute up to $10\%$ to the total halo mass, why its impact
on the critical threshold is below $1\%$?

One can also account for DE energy fluctuations in the density contrast,
defined by
\begin{equation}
\delta_{{\rm tot}}\left(z\right)=\delta_{m}\left(z\right)+\frac{\Omega_{de}\left(z\right)}{\Omega_{m}\left(z\right)}\delta_{de}\left(z\right)\,.\label{eq:total-delta}
\end{equation}
This quantity was also used in nonlinear perturbation theory studies
of CDE by \citet{Sefusatti2011}. When defining the growth function
as 
\begin{equation}
D_{{\rm tot}}\left(z\right)=\frac{\delta_{tot}^{L}\left(z\right)}{\delta_{tot}^{L}\left(0\right)}\,,\label{eq:d-tot-def}
\end{equation}
the change between clustering and homogeneous DE models with the same
background can be about $3-7\%$, while the usual definition with
$\delta_{m}$ only differs about $1\%$ \citet{Batista2017}. Moreover,
this work showed that the $D_{tot}$ and $D_{m}\left(1+\epsilon\right)$,
where $D_{m}$ includes only the matter perturbation, are very similar,
indicating consistency between the linear and nonlinear impact of
CDE when using total quantities.

It is important to note that, due to the DE contribution, the total
mass inside the shell is not conserved during the evolution. After
virialization, however, it has been argued that this contribution
should be stable \citet{Creminelli2010}. In \citet{Basse2012}, this
effect was taken into account in the virial theorem for non-relativistic
particles, yielding the following equation for virialization 
\begin{equation}
\frac{1}{2M_{{\rm tot}}}\frac{d^{2}M_{{\rm tot}}}{dt^{2}}+\frac{2}{M_{{\rm tot}}R}\frac{dM_{{\rm tot}}}{dt}\frac{dR}{dt}+\frac{1}{R^{2}}\left(\frac{dR}{dt}\right)^{2}+\frac{1}{R}\frac{d^{2}R}{dt^{2}}=0\,.\label{eq:basse-virialization}
\end{equation}
Once the equations of the SCM are solved and $\delta^{NL}_{m}\left(z\right)$
and $\delta^{NL}_{de}\left(z\right)$ are known, the virialization time
is determined when (\ref{eq:basse-virialization}) is satisfied. For
further discussion about virialization in the presence of DE, see
\citet{Maor:2005hq}, and for relativistic corrections in the virialization,
\citet{Meyer2012}.

\begin{figure}
\centering{}\includegraphics[scale=0.6]{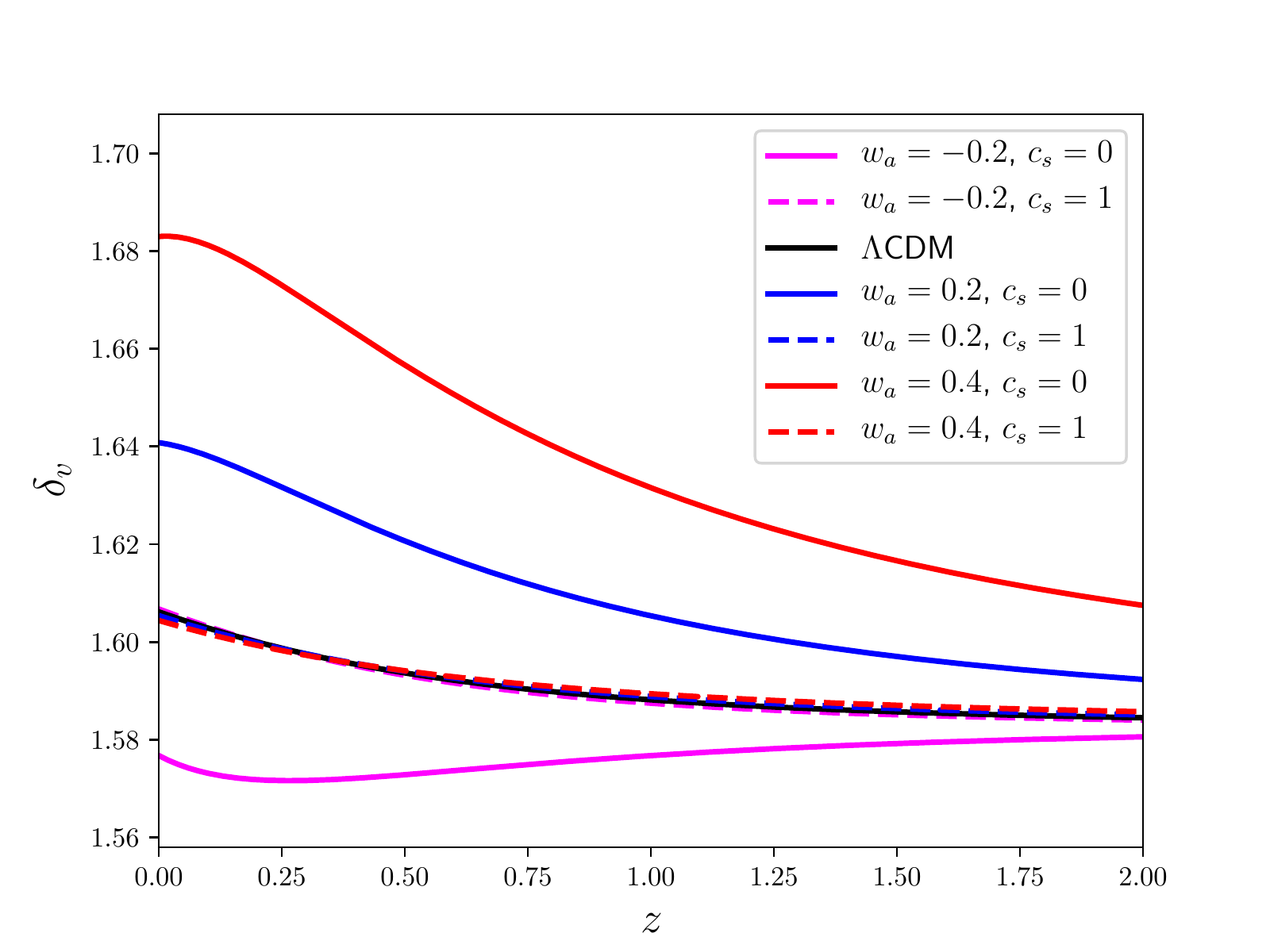}
\caption{Virialization threshold, $\delta_{{\rm v}}$, for homogenous and CDE
models with $w_{0}=-1$, $\Omega_{m0}=0.3$, and different values of $w_{a}$ and $c_{s}$
indicated in the legend. The black solid line represents the $\Lambda$CDM
values and is barely distinguishable from the cases with homogeneousDE.
\label{fig:delta_vir}}
\end{figure}

Having found the virialization redshift, one can compute the virialization
threshold, including the contribution of DE fluctuations

\begin{equation}
\delta_{{\rm v}}\left(z_{v}\right)=\delta_{{\rm tot}}^{L}\left(z_{{\rm v}}\right)\,.\label{eq:d-vir-def}
\end{equation}
As can be seen in figure \ref{fig:delta_vir}, the impact of CDE is
larger than in $\delta_{c}$, reaching up to $4\%$. It is interesting
to note that the magnitude of these differences is more consistent
with the amount of DE fluctuations in halos described by $\varepsilon$
than those observed in $\delta_{c}$.

\begin{figure}
\centering{}\includegraphics[scale=0.6]{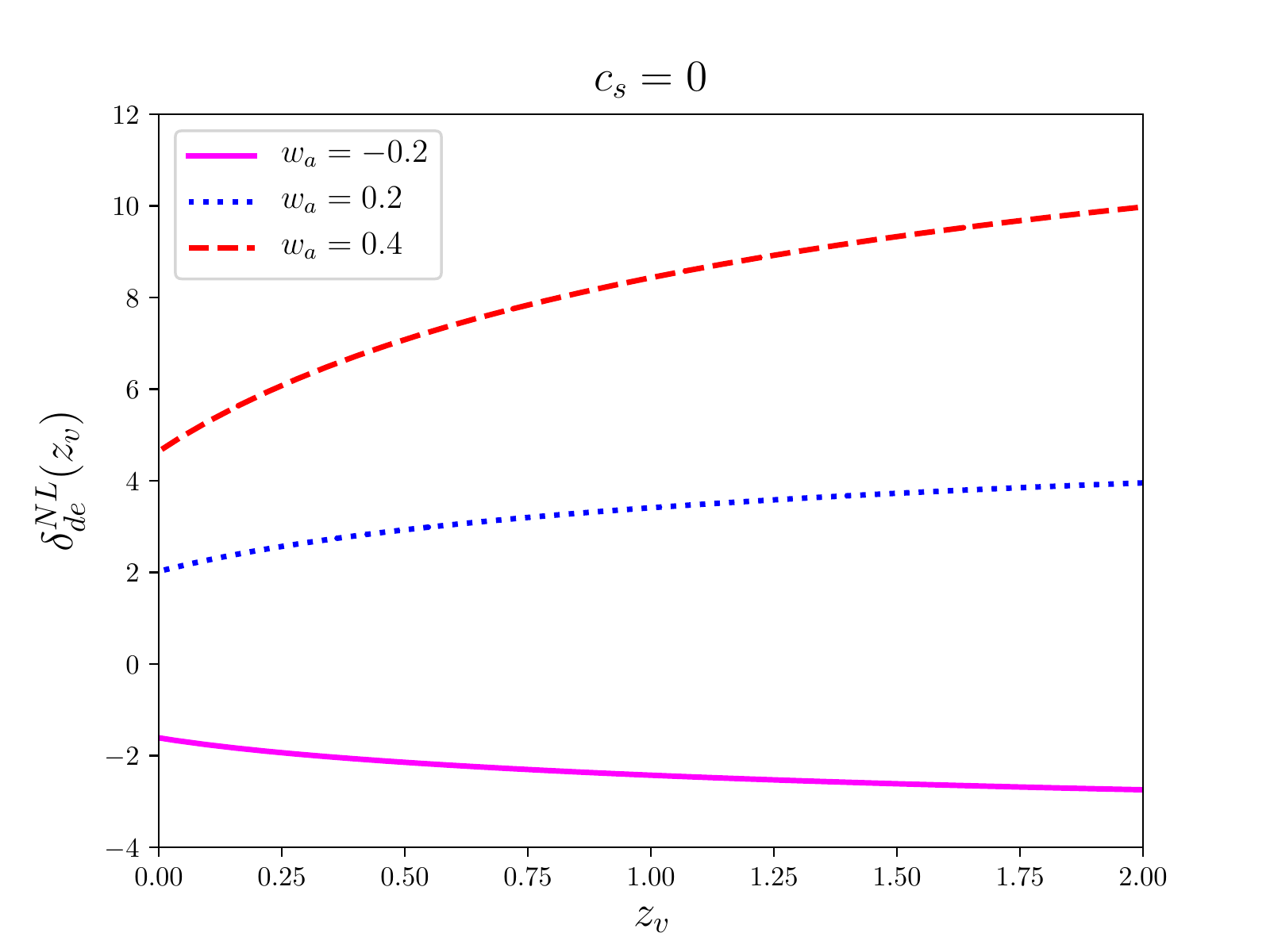}
\caption{Nonlinear DE fluctuation in the clustering case as a function of the virialization redshift for three values of $w_{a}$ shown in the plot legend, $w_{0}=-1$ and $\Omega_{m0}=0.3$
for all models. Note that, for $w_{a}=-0.2$ we have the pathological
values $\delta^{NL}_{de}<-1$.\label{fig:delta_de_vir}}
\end{figure}

It is also important to highlight the behaviour of DE fluctuations.
In figure \ref{fig:delta_de_vir}, we show $\delta_{de}^{NL}\left(z_{v}\right)$.
As can be seen, DE fluctuation can become nonlinear and have a mild
decay at low-$z$, when the EoS of our examples approach $-1$. Clearly,
the larger is $|w_{a}|$, more significant are the DE fluctuations.
We can also see in this plot the pathological case of $\delta^{NL}_{de}<-1$
for phantom DE.

\begin{figure}
\centering{}\includegraphics[scale=0.6]{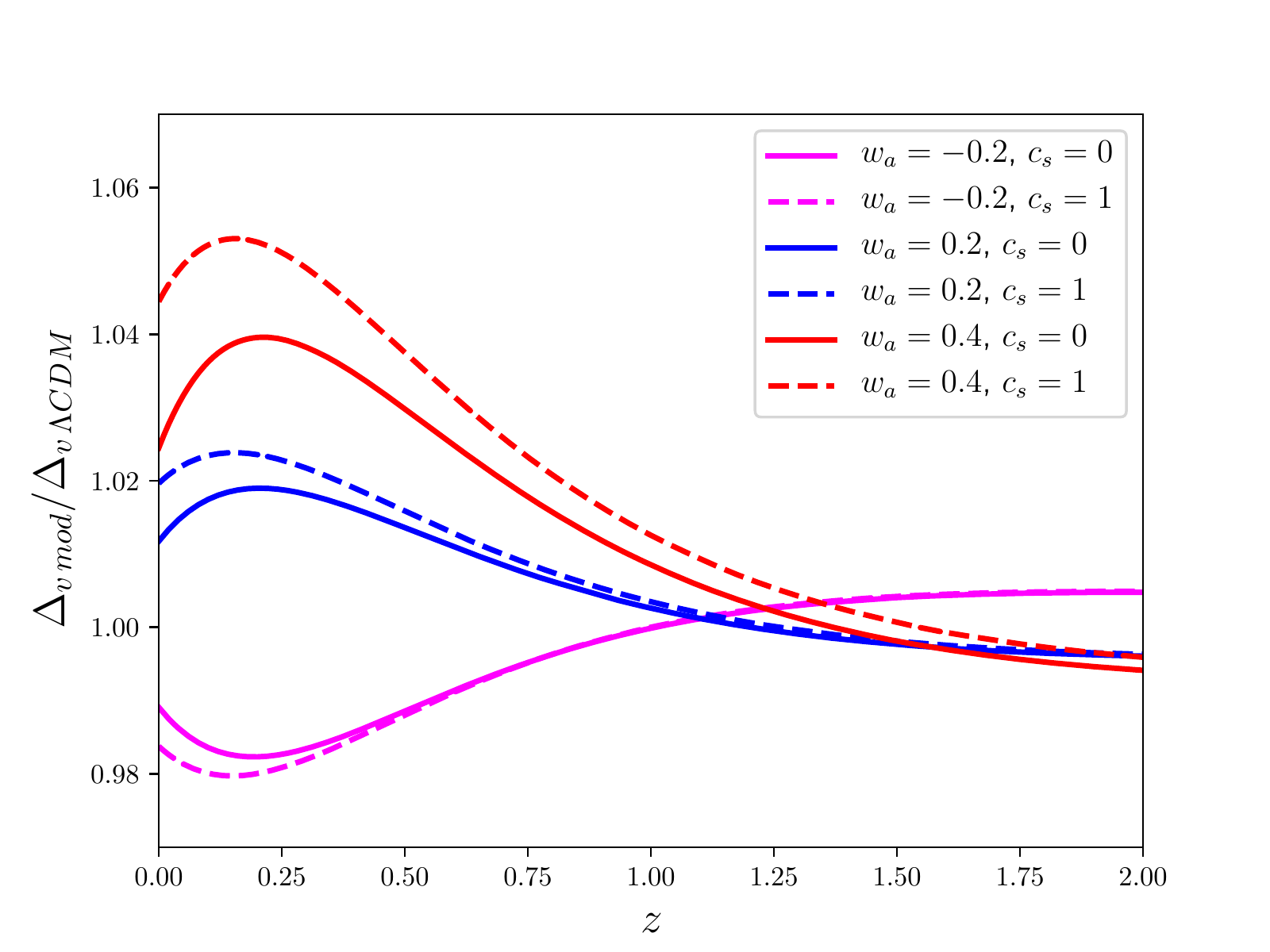}
\caption{Ratio of $\Delta_{{\rm v}}$ of a given model indicated in the legend to the 
corresponding $\Lambda$CDM value. In these examples, $w_{0}=-1$ and $\Omega_{m0}=0.3$
are assumed. 
\label{fig:DVir-variation}}
\end{figure}

In the presence of CDE, the virialization overdensity, defined in
(\ref{eq:Dvir-vir}), is given by
\begin{equation}
\Delta_{{\rm v}}\left(z_{{\rm v}}\right)=\Omega_{m}\left(z_{{\rm v}}\right)\left[1+\delta_{m}^{NL}\left(z_{{\rm v}}\right)\right]+\Omega_{de}\left(z_{{\rm v}}\right)\delta_{de}^{NL}\left(z_{{\rm v}}\right)\,.\label{eq:virial-contrast-clust-de}
\end{equation}
In figure \ref{fig:DVir-variation} we show the relative difference
of $\Delta_{{\rm v}}$ with respect to the $\Lambda$CDM one. Note
that CDE makes $\Delta_{{\rm v}}$ more similar to the $\Lambda$CDM
values.

\section{Halo mass functions\label{sec:Halo-mass-functions}}

Now let us see how to use the findings of the SCM to estimate
the observational impact of DE fluctuations on structure formation.
The computation of the abundance of galaxy clusters relies on the
Halo Mass Function (HMF), which gives the number density of halos
per comoving volume. There are several developments for the HFM, including
an analytic model based on the spherical collapse by \citet{Press:1973iz} and
a semi-analytic approach based on ellipsoidal collapse \citet{Sheth1999}.
More recently, N-body numerical simulations were used to determine
fitting functions for the HMF, e.g., \citet{Warren2006,Tinker2008b}. Besides determining the abundance of galaxy clusters, the HMF can be used to compute the nonlinear power spectrum, \citet{Cooray2002}.

The proper approach to study the impact of CDE on the abundance of
galaxy clusters would be to include its dynamical effects in the numerical simulations
of structure formation. Codes with
this capability have been developed only quite recently, but without
determining the associated HMF. Let us first discuss developments
based on analytic and semi-analytic HMF and then some relevant results
from simulations. 

We anticipate that there is no consensus of how to implement the impact of CDE
in HMF. As a consequence, the predictions also vary, and the
dection of DE fluctuations also depend on the particular implementation used to include the effects of CDE on halo abundances.

The Press-Schechter (PS) HMF \citet{Press:1973iz} assumes that the
matter density field, $\delta$, smoothed on some scale, has a Gaussian probability
distribution
\begin{equation}
p\left(\delta\right)=\frac{1}{\sqrt{2\pi}\sigma}\exp\left(-\frac{\delta^{2}}{2\sigma^{2}}\right)\,,\label{eq:pdf-delta}
\end{equation}
The variance of the smoothed field, is given by
\begin{equation}
\sigma^{2}\left(R\right)=\int\frac{dk}{k}\frac{k^{3}P\left(k\right)}{2\pi^{2}}|W\left(kR\right)|\,,\label{eq:variance}
\end{equation}
where $P\left(k\right)=|\delta_{k}|^{2}$ is the linear matter power
spectrum,
\begin{equation}
W\left(kR\right)=\frac{3}{\left(kR\right)^{3}}\left[\sin\left(kR\right)-kR\cos\left(kR\right)\right]\label{eq:top-hat-window}
\end{equation}
is a top-hat window function with mass-scale relation given by
\begin{equation}
R=\left(\frac{3M}{4\pi\bar{\rho}{}_{m0}}\right)^{1/3}\label{eq:mass-scale relation}
\end{equation}
 and $\bar{\rho}{}_{m0}$ the background matter density now.

Assuming that regions with $\delta>\delta_{c}$ form bound structures,
where $\delta_{c}$ is a certain threshold value to be defined, the
fraction of such objects with mass greater than $M$ is given by
\begin{equation}
P\left(\delta>\delta_{c}\right)=\int_{\delta_{c}}^{\infty}p\left(\delta\right)d\sigma=\frac{1}{2}\left[1-\text{erf}\left(\frac{\delta_{c}}{\sqrt{2}\sigma}\right)\right]\,.\label{eq:prob-delta}
\end{equation}
The comoving number density of halos per mass interval is then given
by
\begin{equation}
\frac{dn\left(M,z\right)}{dM}=\frac{\bar{\rho}{}_{m0}}{M}f_{PS}\left(M,z\right)\,,\label{eq:comving-density}
\end{equation}
where 
\begin{equation}
f_{PS}\left(M,z\right)\equiv2\left|\frac{\partial P}{\partial M}\right|\,\label{eq:PS-multiplocity}
\end{equation}
is the PS multiplicity function. The factor $2$ was first introduced
in an ad hoc manner, to guarantee the normalization of the HMF \citet{Press:1973iz},
\begin{equation}
\int_{0}^{\infty}\frac{dn\left(M,z\right)}{dM}dM=1\label{eq:normalization}
\end{equation}
but it can be formally determined \citet{1990MNRAS.243..133P,Bond:1990iw}.

Assuming that only $\sigma$ is mass-dependent, we have the usual
form of the PS multiplicity function:
\begin{equation}
f_{PS}\left(M,z\right)=\sqrt{\frac{2}{\pi}}\frac{1}{M}\frac{\delta_{c}\left(z\right)}{\sigma\left(M,z\right)}\left|\frac{\partial\ln\sigma\left(M,z\right)}{\partial\ln M}\right|\exp\left(-\frac{\delta_{c}^{2}\left(z\right)}{2\sigma^{2}\left(M,z\right)}\right)\,.
\end{equation}

In the first studies with the PS-HMF, the values of $\delta_{c}$
assumed varied in the range $\left(1,10\right)$, e.g., \citet{Press:1973iz,10.1093/mnras/189.2.203,Colafrancesco1989,Gelb1994}.
Later, it became usual to use the constant EdS collapse threshold
value $\delta_{c}\simeq1.69$, e.g., \citet{1988MNRAS.231P..97N,1990MNRAS.243..133P}.
In the presence of $\Lambda$ (for open, flat and closed models) \citet{1992ApJ...386L..33L}
reported $1.64<\delta_{c}<1.73$ at low-$z$. As we saw in subsection
(\ref{subsec:SC-model-homo}), other works also determined fitting
functions for these quantities for homogeneous DE models \citet{Kitayama1996,Weinberg2003},
also finding a small deviation from the EdS value, of less than $1\%$.

The matter power spectrum can be numerically determined by codes like
CAMB \citet{Lewis2000} or CLASS \citet{Lesgourgues2011}. It can
also be given by fitting functions, which, in most cases, can be separated
in the following form
\begin{equation}
P\left(k\right)=P_{p}\left(k\right)T^{2}\left(k\right)D_{m}^{2}\left(z\right)\,,\label{eq:PS-fitting}
\end{equation}
where $P_{p}\left(k\right)$ is the primordial power spectrum associated
to matter perturbations given by the inflationary model, $T\left(k\right)$
is the transfer function (for instance, given by \citet{Eisenstein1998})
and 
\begin{equation}
D_{m}\left(z\right)=\frac{\delta_{m}\left(z\right)}{\delta_{m}\left(z=0\right)}
\end{equation}
the linear growth function of matter perturbations. Since the amount
of DE is very small around the recombination time, the transfer function
is essentially unaffected by DE. The main impact of DE, either smooth
or inhomogeneous, is on the growth function, which strongly depends
on the cosmological evolution at low-$z$.

In this context, for homogeneous DE models, the only relevant modification
on the HMF occurs via the modifications on the growth function caused
by different evolutions of $w$. Then it is expected that either analytic
or numerical HMF can directly incorporate the impact of DE via the
modifications of the growth function. Several works have studied homogeneous
DE in such scenario, either with analytic approaches \citet{Percival:2005vm,LeDelliou:2005ig,Horellou:2005qc,Liberato:2006un,Bartelmann2006,Pace2010a,Pace:2011kb}
or numerical simulations \citet{Linder:2003dr,Grossi2009}.

For DE models with arbitrary $c_{s}$, both $\delta_{c}$ and $D_{m}$
become mass-dependent because DE fluctuations are enhanced above the
sound horizon scale, and impact matter perturbations in a scale-dependent
manner. This case is studied in \Citet{Basse2011,Basse2012,Basse2014}.
Given that $\delta_{c}\left(z\right)\rightarrow\delta_{c}\left(M,z\right)$,
the PS multiplicity function acquires an extra mass-dependent term 
\begin{equation}
f_{PS}\left(M,z\right)=\sqrt{\frac{2}{\pi}}\frac{1}{M}\frac{\delta_{c}\left(M,z\right)}{\sigma\left(M,z\right)}\left|\frac{\partial\ln\sigma\left(M,z\right)}{\partial\ln M}-\frac{\partial\ln\delta_{c}\left(M,z\right)}{\partial\ln M}\right|\exp\left(-\frac{\delta_{c}^{2}\left(M,z\right)}{2\sigma^{2}\left(M,z\right)}\right)\,.\label{eq:PS-HMF-delta-m}
\end{equation}
Although this new mass dependence is a feature of CDE if $c_{s}$ is
associated with some mass scale that can be probed by the observed
abundance of galaxy clusters, it is much smaller than the mass dependence
on $\sigma$.

In the limit of CDE, $c_{s}\rightarrow0$, the growth of $\delta_{de}$
has the same scale dependence of matter perturbations. Then both $\delta_{c}$
and $D_{m}$ remain scale independent. In this case, the scale
dependent feature due to CDE in the HMF vanishes, but its impact can
be larger. Several papers have studied this scenario: \citet{Nunes:2004wn,Manera:2005ct,Abramo2007,Abramo2009a,Creminelli2010,Batista:2013oca,Pace2014a,Batista2017,Heneka2018}.

Now let us consider the Sheth-Tormen HMF \citet{Sheth1999} and how
it can be modified to include the effects of DE fluctuations. Since
it provides a better description of cluster abundances given by numerical simulations,
we expect it to provide better estimates of the impact of DE fluctuations on
clusters abundances. The original ST-HMF is given by
\begin{equation}
\frac{dn_{ST}}{dM}=-A\sqrt{\frac{2a}{\pi}}\left[1+\left(\frac{a\delta_{c}^{2}\left(z\right)}{\sigma^{2}\left(M,z\right)}\right)^{-p}\right]\frac{\rho_{m0}}{M^{2}}\frac{\delta_{c}\left(z\right)}{\sigma\left(M,z\right)}\frac{\partial\ln\sigma\left(M,z\right)}{\partial\ln M}\exp\left(-\frac{a\delta_{c}^{2}\left(z\right)}{2\sigma^{2}\left(M,z\right)}\right)
\end{equation}
$A=0.2162$, $p=0.3$ and $a=0.707$. As in the case of PS theory,
it is usual to use $\delta_{c}\simeq1.69$. Interestingly, in the
context of EdS model, the parameter $a$ reduces the effective
threshold $\delta_{c\,{\rm eff}}=\sqrt{a}\delta_{c}\simeq1.42$, indicating
that $\delta_{{\rm v}}\simeq1.52$, (\ref{eq:critical-delta-virial}),
gives a better description than $\delta_{c}\simeq1.69$. This fact
can be interpreted as an indication that the spherical collapse quantities defined
at virialization time provide a more natural description of the threshold
and virialization density.

\citet{Despali2016} have used several overdensities criteria to detect
halos in simulations and fit the three parameters of ST-HMF. It was
found that when halos are identified with the viral overdensity $\Delta_{cc}$,
defined in (\ref{eq:Dvir-coll-crit}), the ST-HMF is nearly universal
with respect to redshift and $\Omega_{m0}$ variations. These parameters
can also be constrained by future observations, see \citet{Castro2016}.

Following these ideas, \citet{Batista2017} proposed that the proper
threshold density for halo formation in the presence of CDE should
be modified to include the contribution of DE fluctuations at virialization time,
and used virial threshold, (\ref{eq:critical-delta-virial}). But, in order
to make a more conservative estimate, the usual parameter $a$ was
rescaled by $a\rightarrow\tilde{a}=a\delta_{c}^{2}/\delta_{{\rm v}}^{2}$.
With this choice, the ST-HMF is unchanged in the EdS limit. In figure
\ref{fig:delta_vir}, the evolution of $\delta_{{\rm v}}$ is shown
for some homogenous and CDE models.

\begin{figure}
\centering{}\includegraphics[scale=0.45]{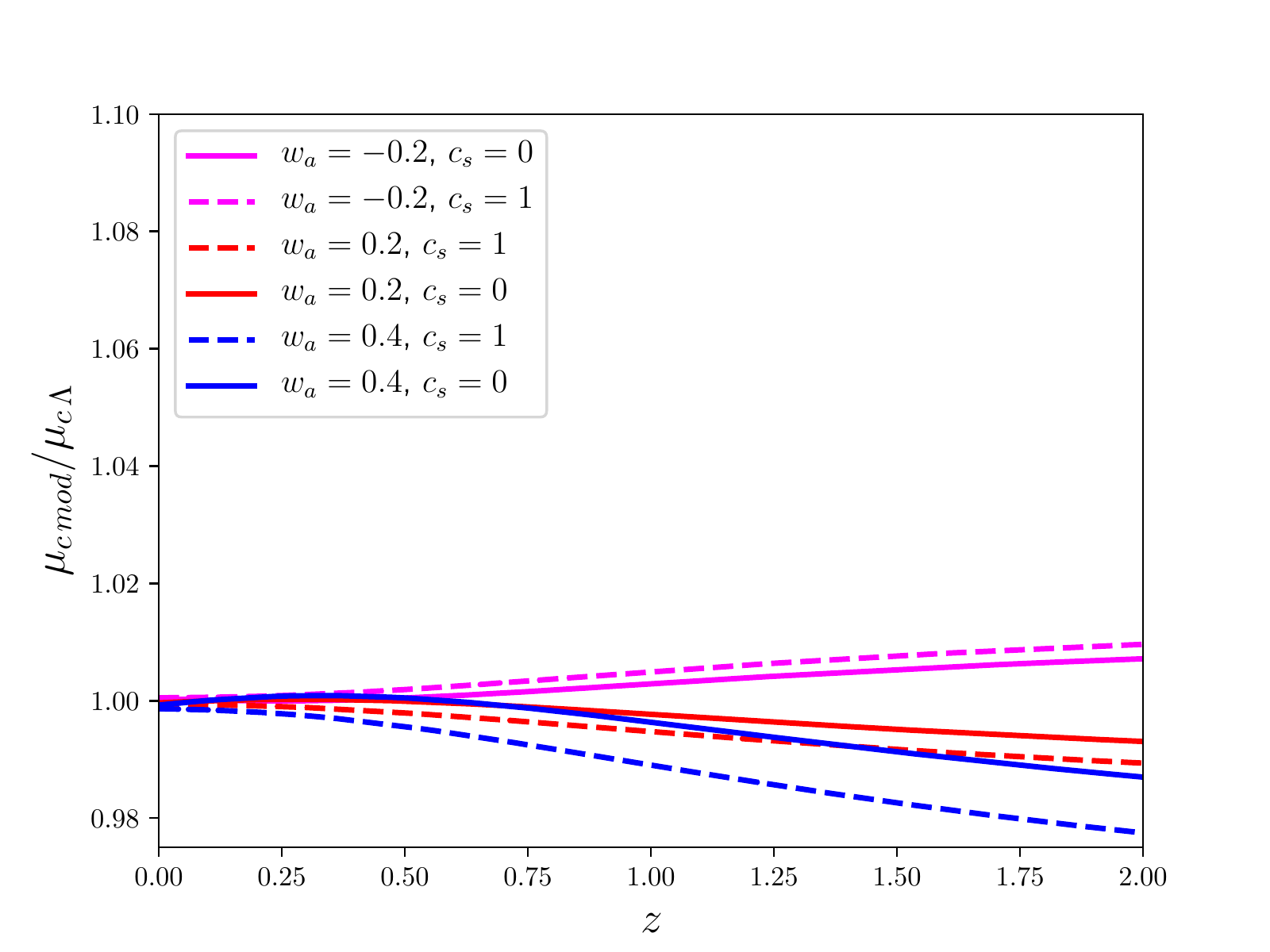}\includegraphics[scale=0.45]{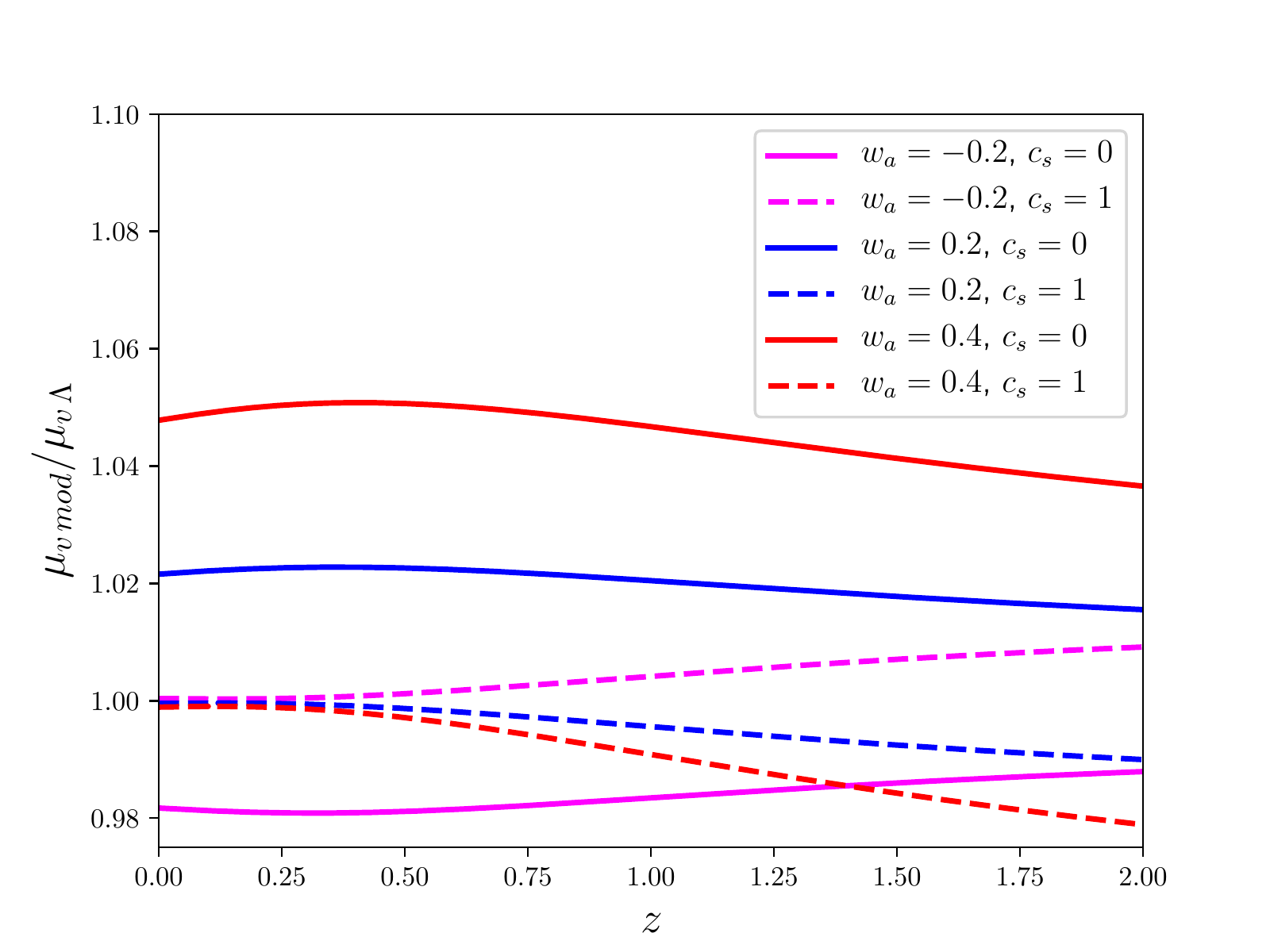}\caption{Evolution of $\mu_{c}$ (left panel) and $\mu_{{\rm v}}$ (right panel)
with redshift, defined in equation (\ref{eq:defs-nus}), divided by
the corresponding function in the $\Lambda$CDM model for homogenous
and CDE models with $w_{0}=-1$ and different values $w_{a}$ and
$c_{s}$ indicated in the legends. Due to the small impact of distinct
DE models on $\delta_{c}$, the differences in $\mu_{c}$ with respect
to the case with $\Lambda$ is very small at $z\simeq0$, but can
reach about $2\%$ at $z\simeq2$. In the case of CDE, the differences
in $\mu_{{\rm v}}$ are larger and also present at low $z$. \label{fig:nus-diff}}
\end{figure}

In practice, for a given mass scale, the abundance of massive halos
strongly depends on the quantities
\begin{equation}
\mu_{c}=\frac{\delta_{c}\left(z\right)}{D_{{\rm m}}\left(z\right)}\text{ and }\mu_{{\rm v}}=\frac{\delta_{{\rm v}}\left(z\right)}{D_{{\rm tot}}\left(z\right)}\,.\label{eq:defs-nus}
\end{equation}
In figure \ref{fig:nus-diff} we show the ratio of these functions
with respect to the corresponding values in the $\Lambda$CDM model.
The impact of CDE is larger in $\mu_{{\rm v}}$ and is nearly constant
with redshift-dependent. Having in mind that, although $\delta_{de}\left(z\right)$
decays at lower $z$, when $w\rightarrow-1$, the quantity $\delta_{de}\left(z\right)\Omega_{de}\left(z\right)/\Omega_{m}\left(z\right)$,
which impacts both $\delta_{{\rm v}}$ and $D_{tot}$, is nearly constant.
Thus, the behaviour of $\nu_{{\rm v}}$ seems more consistent with
the effective contribution of DE fluctuation in the dynamics of the
collapse shown in figure (\ref{fig:delta_de_vir}).

Another important point regarding the impact of CDE on HMFs is that
they are not calibrated by numerical simulations. A possible approach
to this issue, also used in the context of non-gaussianities \citet{LoVerde2008}
and baryonic feedback \citet{Velliscig2014}, is to multiply the more
accurate numerically calibrated HMF by a factor that encodes the relative
impact of DE fluctuations with respect to the usual homogeneous model
\citet{Creminelli2010,Batista2017,Heneka2018}
\begin{equation}
\frac{\left(dn/dM\right)_{c_{s}=0}}{\left(dn/dM\right)_{c_{s}=1}}\,.\label{eq:de-HMF-correction}
\end{equation}

Moreover, one has to take into account how CDE affects the cluster
mass. It is expected that the contribution of DE to mass shifts HMF
according to $M\rightarrow M\left(1-\epsilon\right)$ \citet{Creminelli2010,Batista:2013oca,Batista2017}.
\citet{Batista2017} also analyzed modifications on the mass-scale
relation, equation (\ref{eq:mass-scale relation}), and normalization
condition. Whereas the former can be safely neglected, the latter
is of the order of $\epsilon$, possibly modifying HMF about a few
per cent on all mass scales. These studies have shown that, depending
on the evolution of $w$, the abundance of massive halos can change
by $10-30\%$ with respect to homogeneous DE models. But this change
can be even larger very massive halos, $M>10^{15}M_{\odot}$, whose abundance
is very sensitive to modifications on the exponential tail of HMF.

\subsection*{Numerical simulations}

Recently, numerical simulations capable of treating CDE were developed.
The first approach proposed was to include DE linear perturbations
given by Einstein-Boltzmann solvers as a new source of the gravitational
potential in N-Body simulations \citet{Dakin2019}. Of course, this
implementation can not deal with nonlinear DE fluctuations, but was
crucial in confirming that linear and mildly nonlinear DE perturbations
have a non-negligible impact on structure formation.

\citet{Hassani2019} developed a code capable of describing nonlinear
DE fluctuations. They showed that models with $w=-0.9$ and $c_{s}=10^{-3.5}$
do impact matter fluctuations and the gravitational potential on small
scales. They also found that DE and matter fluctuations are correlated,
as indicates the solution (\ref{eq:delta-de-null-cs}). However, HMF
was not computed. \citet{Hassani2020} also studied the imprint of
CDE on observables associated with the gravitational potential. They
confirmed the tendency of CDE to compensate for the changes in the
background and found modifications of $2-5\%$ on the observables
studied. It is important to note that, due to the choice $w=-0.9$,
the imprints of DE fluctuations found in these works are not as large
as in models in which $w$ is smaller at intermediate
redshifts.

Although these numerical studies have confirmed several results from
perturbation theory and the SCM, the distinct proposals to implement
the impact of CDE on HMF were not yet tested by simulations. Unfortunately the
results vary between these implementations,
and there is no definitive prediction about the actual impact of CDE
on the abundance of galaxy clusters. See also the discussion in \citet{Basse2014}

\section{Cosmological observables}

As we saw, DE fluctuations can potentially impact the linear and nonlinear
evolution of matter fluctuations and the gravitational potential.
Thus, many observables like CMB anisotropies, linear and nonlinear
matter power spectrum and growth rate can change due to the presence
of a new clustering component. We have already discussed some of those
effects, especially regarding HMF. Next, overview of some
observational strategies discussed in the literature that can possibly
detect CDE. The grouping of observables shown below is somewhat arbitrary
because most of the works presented discuss and analyze combinations
of them.

\subsection{CMB and Large Scale Structure}

\citet{Weller2003} attempted to constrain the value of $c_{s}$ using
the first WMAP data release and found that $c_{s}$ is unconstrained.
In a similar analysis, \citet{Bean2004}
reported a $1\sigma$ constraint $c_{s}<0.2$. Later, \citet{Hannestad2005}
included Large Scale Structure data in the analysis, showing that
$c_{s}$ remains essentially unconstrained. \citet{Putter2010} also reached
the same conclusion, but showing that chances of detection of CDE
are larger in Early DE models.

\citet{Takada:2006xs} showed that a $2000$$\text{deg}^{2}$ galaxy
redshift survey at $z\simeq1$, together with CMB information from
Planck, can distinguish between smooth ($c_{s}=1$) and CDE with $c_{s}<0.02$
and $w=-0.95$.

Analyzing Early DE models and using CMB and LSS data, \citet{Bhattacharyya2019}
reported $c_{s}=0.37$ and $\Omega_{de}\left(a_{rec}\right)=0.02$.
It was shown that, when $c_{s}$ is allowed to vary, the contribution
of DE in the early Universe can be more significant.

It has also been shown that the cross-correlation between galaxy survey
and the Integrated Sachs-Wolf effect is a promising technique to detect
$c_{s}$ \citet{Hu2004}. This idea was further explored in \citet{Corasaniti2005,Pietrobon2006,Ballesteros2010b,Li2010}.

The most recent analysis of CDE by the Planck team \citet{Ade2016}
indicates that $c_{s}$ is unconstrained. However, $w$ was assumed
constant. Since at late times we must have $w\simeq-1$, DE fluctuations
are very small in this scenario, and the impact of $c_{s}$ on observables
is negligible. Only models with time-varying $w$ can present relevant
DE fluctuations, like in the case o Early DE.

\subsection{Higher order perturbation theory}

CDE has also been studied in the framework of higher-order
perturbation theory by \citet{Sefusatti2011,DAmico2011,Anselmi2011,Anselmi2014}.
These works found an impact of a few per cent on nonlinear corrections to the
linear power spectrum. These authors also understand that, in CDE
models, the total perturbation is the relevant one to be considered,
not only matter perturbations.

\subsection{Weak lensing}

\citet{Sapone2010} investigated how tomographic weak lensing and
galaxy redshift surveys can constrain DE sound speed. Considering
$w=-0.8$ they found that, if $c_{s}<0.01$, the sound speed can be
constrained. More studies in this direction include \citet{Ayaita2012,Majerotto2016}

\subsection{Cluster abundances}

\citet{Abramo2009a} have forecasted the constraining power of galaxy
clusters surveys on $c_{s}$. Although the implementation used is
not entirely consistent with SCM in the sense of how $c_{s}$ is varied,
what artificially enhances the dependence of cluster abundances on
this parameter, some interesting conclusions were drawn. It was shown
that future experiments like Euclid can play a decisive role in detecting
DE fluctuations and that they impact the constraints on $w_{0}$ and
$w_{a}$ at $10-30\%$ level.

\citet{Basse2014} has also forecasted how $c_s$ can be constrained by data from 
cluster abundances, CMB, cosmic shear and galaxy clustering correlations. It was found that,
considering $w_0=-0.83$ and $w_a=0$, future Euclid data can distinguish between $c_s=1$ and $c_s=0$. However, this result strongly depends on how the impact of CDE is implemented on HMF. In particular, it was found that, when considering the DE contribution to the cluster mass, the sensitivity on $c_s$ significantly degrades. 

\citet{Appleby2013a} showed that, using the Euclid satellite cluster
survey together with Planck data, Early DE models with $c_{s}<0.1$
and $\Omega_{de}\left(a_{{\rm rec}}\right)=0.009$ can be distinguished
from models with $c_{s}=1$. In such models, DE is not negligible
at high-$z$ and the EoS parameter can be larger than $-1$ at high-$z$.
Hence, the impact of DE perturbations is strongly enhanced. The authors
noted that $c_{s}$ is mainly constrained by CMB data. Although this
work used the Tinker HMF \citet{Tinker2008b}, which does not take
into account the nonlinear effects of DE fluctuations, the reported
insensitiveness of cluster abundance on clustering Early DE model
is in accordance with \citet{Batista:2013oca}. The reason for this is
the following: whereas Early DE has a strong impact on the background
evolution, decreasing the matter growth, if it clusters, the perturbations
partially compensate for this change. Note, however, that this result
was obtained using $\delta_{c}$ and $D_{m}$. When using $\delta_{{\rm v}}$
and $D_{{\rm tot}}$ in HMF, it is expected that more pronounced changes
due to CDE will be present.

\citet{Heneka2018} used cluster data from several experiments plus
CMB, Barion Acoustic Oscillations (BAO) and Supernova Ia (SNIa) observations to constraint cosmological parameters with $c_{s}=1$, $c_{s}=0$ and constant $w$. It was shown that the allowed
regions in the $\sigma_{8}-\Omega_{m0}$ plane change due to CDE.
It was found that $w<-1$ is preferred by data and that $\sigma_{8}$
is reduced when $c_{s}=0$. This impact of CDE can alleviate the tension
in cluster data found by Planck \citet{Ade2015q}.

It is important to note that, a possible issue regarding constraints
of CDE models with cluster data is associated with the observable-mass
scaling relations, see \citet{Mantz2010,Kravtsov2012}. So far, no
analytical or numerical study has been conducted to explore the impact
of DE fluctuation on these relations. As discussed before, DE fluctuations
certainly impact the gravitational potential, thus lensing signals,
X-ray can SZ luminosities relations with the cluster mass can be affected.

\subsection{Internal structure of galaxy clusters}

The impact of CDE on the internal structure of halos was studied in
\citet{Mota:2008ne,Basilakos2009c}. As it would be expected from the
results that $R_{{\rm ta}}/R_{{\rm v}}<0.5$ in the presence of DE
and its possible perturbations, the concentration parameter of galaxies
clusters increases. \citet{Basilakos2009c} have analyzed the mass-concentration
relation of four massive clusters, showing that the data is better described by CDE models.

\subsection{$S_{8}$ tension and growth rate}

Clustering DE models can be used to reduce the value of $\sigma_{8}$,
\citet{Kunz2015}. The parameter $S_{8}=\sigma_{8}\sqrt{\Omega_{m0}/0.3}$
inferred by CMB observations using the $\Lambda$CDM model, \citet{Aghanim2020},
is about $3\sigma$ in tension with the values determined by weak
lensing observations \citet{Asgari2021,Abbott2021}, see also \citet{DiValentino2021}
and references therein. To explore the impact of CDE -- here we will
focus on the variation of $\sigma_{8}$ only -- we fix $\sigma_{8mod}\left(z\right)=\sigma_{8mod}D_{mod}\left(z\right)$,
for a given model "$mod$", to have the same value as in the $\Lambda$CDM
model at $z_{rec}$
\begin{equation}
\sigma_{8mod}=\frac{D_{\Lambda}\left(z_{rec}\right)}{D_{mod}\left(z_{rec}\right)}\sigma_{8\Lambda}\,,
\end{equation}
where $\sigma_{8\Lambda}=0.8111$ for $\Omega_{m0}=0.3153$ \citet{Aghanim2020}.

\begin{figure}
\centering{}\includegraphics[scale=0.45]{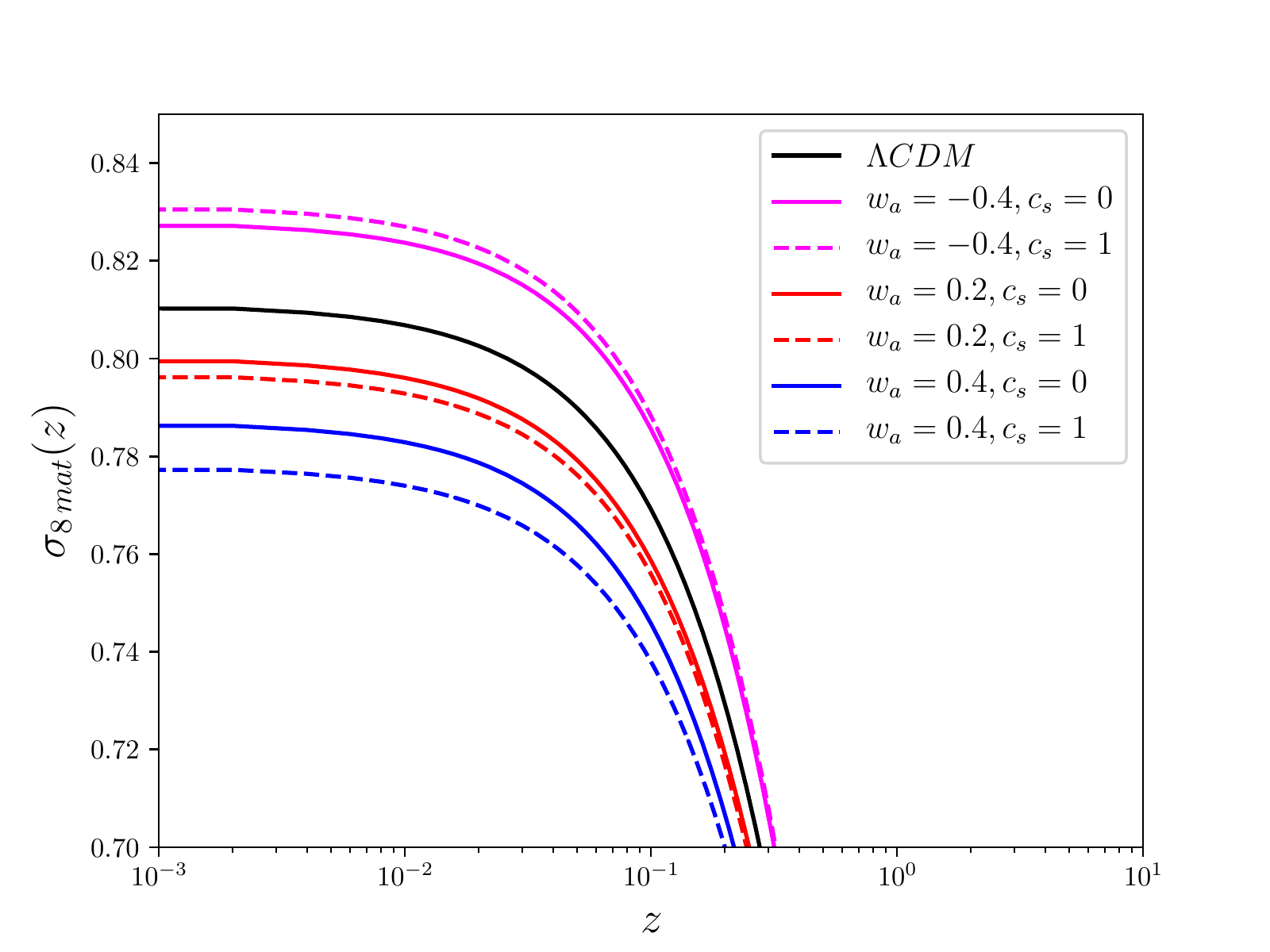}\includegraphics[scale=0.45]{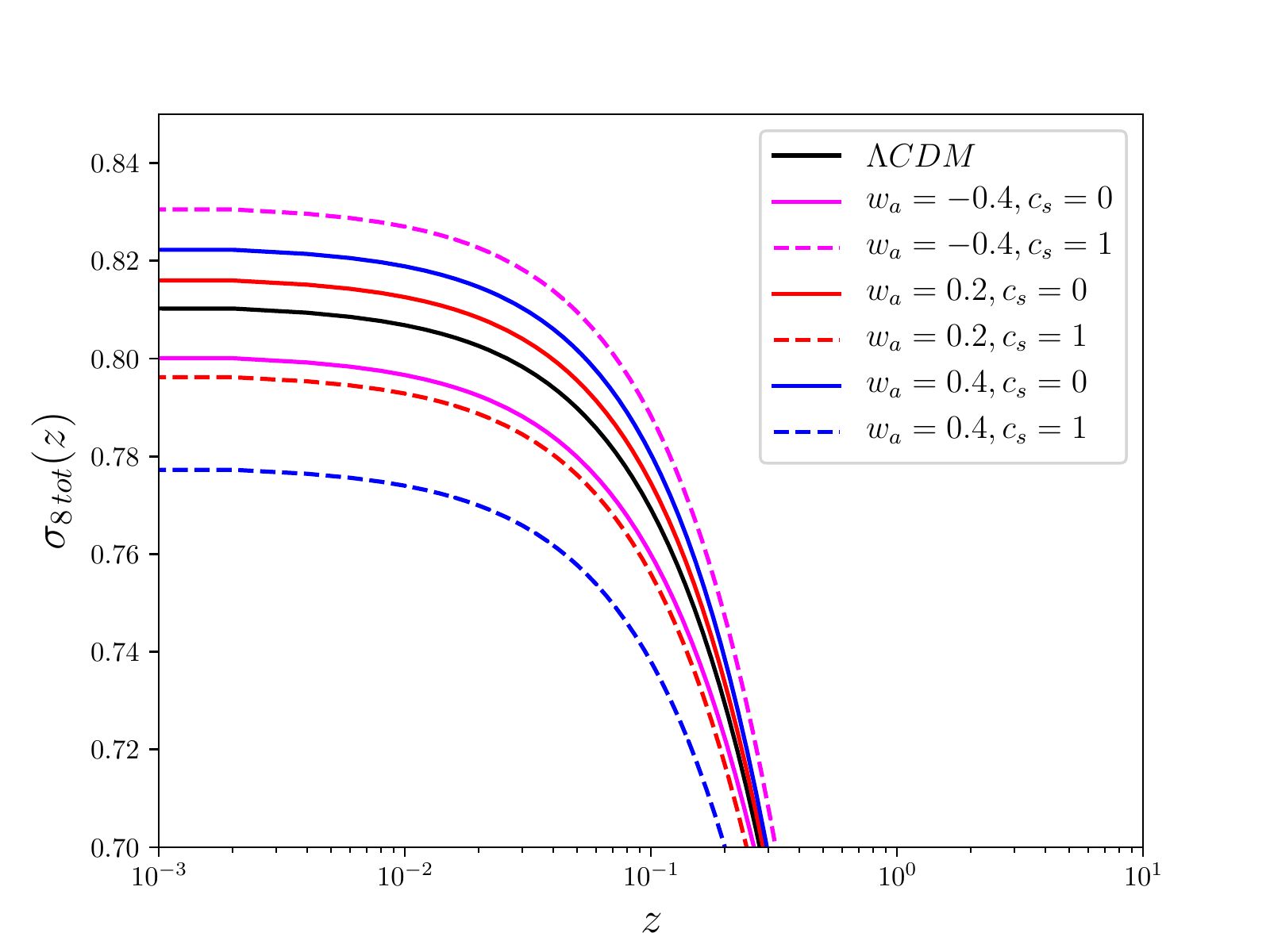}\caption{Redshift evolution of $\sigma_{8}\left(z\right)$ for various DE models
indicated in the legends. In the left panel: the growth function is
given only by the matter perturbation. Right panel: the growth function
is computed including the CDE contribution given by (\ref{eq:total-delta}).\label{fig:sigma8}}
\end{figure}

As can be seen in the left panel of figure (\ref{fig:sigma8}), in
the case where only matter perturbations contribute to the growth
function, CDE make the evolution of $\sigma_{8}\left(z\right)$ more
similar to $\Lambda$CDM. In the right panel, when DE perturbations
are included in the growth function, we see that the impact of CDE
is much larger. In this case, for $w_{a}>0$ ($w_{a}<0$), DE perturbations
can make $\sigma_{8}\left(z\right)$ larger (smaller) than in $\Lambda$CDM.

Homogeneous phantom DE increase $\sigma_{8}$ because $\Omega_{de}\left(z\right)$
grows rapidly at low-$z$, delaying the suppression of matter perturbations
when compared to $\Lambda$CDM. When DE perturbations are allowed,
given that $\delta_{de}<0$, matter growth is suppressed. Considering
the total growth function, this effect can potentially alleviate the
$S_{8}$ tension.

On the other hand, homogeneous non-phantom DE decrease $\sigma_{8}$
because $\Omega_{de}\left(z\right)$ starts to grow earlier than in
$\Lambda$CDM model. Clustering DE now partially compensates for this
decay in the growth of matter perturbations, enhancing $\sigma_{8}$.
For the case of the total growth function, CDE can potentially worsen
the $S_{8}$ tension.

An analysis of CDE using cluster, CMB, BAO and SNIa data was done
by \citet{Heneka2018}. It was found the $\sigma_{8}$ is reduced
for $c_{s}=0$ with slight increase in $\Omega_{m0}$. Considering
the best fit values, the overall effect is a reduction of $S_{8}$
in CDE models. The results also show that, assuming constant $w$,
phantom values are prefered by data.

Clustering DE also affects the growth rate of matter \citet{Batista2014a,Mehrabi2015,Mehrabi2015a},
given by
\begin{equation}
f\left(z\right)=\frac{d\ln\delta_{m}}{d\ln a}\,
\end{equation}
In particular, \citet{Mehrabi2015b} reported that CDE models can
fit $f\left(z\right)\sigma_{8}\left(z\right)$ data better than $\Lambda$CDM.

\section{Discussion}

The variety of models that try to explain the cosmic acceleration
is enormous. The most studied and tested ones either do not present
DE fluctuations ($\Lambda)$ or have negligible fluctuations on scales
well below the horizon (quintessence). However, many other models
can present a low $c_{s}$ value and relevant DE fluctuations on small
scales. However, the actual amount of DE fluctuations also depends
on the evolution of EoS. If $w\simeq-1$ throughout the cosmic evolution,
DE fluctuations are very small, regardless of the value of $c_{s}$.

Hence, the prospects to detect DE fluctuations on small scales, which
would rule out $\Lambda$ and quintessence as possible drivers of
cosmic acceleration, strongly depend on how far from $-1$ the EoS
can be in the past. Although $\Lambda$ is in excellent agreement
with almost all cosmological data available, there are no strong constraints
on how much $w$ can deviate from $-1$ at intermediate redshifts.
For instance, the allowed parameter space for $w_{0}$ and $w_{a}$
for the case of $c_{s}=1$ is still very large \citet{Ade2016}. As we reviewed, for constant 
EoS, DE models with $w\simeq -0.9$ and $c_s<10^{-2}$ leave a detectable impact
on several cosmological observables. Being more conservative with the value of $w$ now, we showed that, even with $w_{0}=-1$, DE fluctuations become nonlinear when assuming $|w_a|>0.2$ and $c_s=0$, and
influence structure formation.

The linear evolution and the corresponding impact of DE perturbations
is well understood and implemented in numerical codes like CAMB and
CLASS. The nonlinear evolution in the presence of CDE is, however,
more complicated and much less developed. First efforts in this direction
used the SCM and only recently numerical codes for the nonlinear evolution
were developed. The early findings about the phenomenology of CDE
are in agreement with those found by recent numerical results. These
studies have shown that DE fluctuations can become nonlinear and do
impact matter fluctuations, the formation of halos and the gravitational
potential.

The abundance of massive galaxy clusters is certainly one of the most
affected observables. However, up to now, no numerical simulations
modelled the impact of CDE in HMF. The analytical motivated proposals
for HMF in the presence of CDM are divided into two main groups: one
in which the impact of CDE enters only via the matter quantities ($\delta_{c}$,
$D_{m}$, $\sigma_{8m}$) and another based on the total weighted
fluctuations ($\delta_{{\rm v}}$, $D_{{\rm tot}}$, $\sigma_{8{\rm tot}}$).
Naturally, the impact of CDE is enhanced in the latter type of modifications.
Moreover, these different recipes can be combined with mass rescalings
due to DE contribution, $M\rightarrow M\left(1+\epsilon\right)$. However, there is
no consensus on the literature on which is the accurate implementation to compute the 
effect of CDE on halo abundances.

An important theoretical issue with CDE models is associated with
phantom models. For $c_{s}\rightarrow0$, one can have $\delta_{de}<-1$
in the nonlinear regime. At first glance, this indicates that these
models are inconsistent. A possible realization of phantom clustering
models may include imperfect fluids \citet{Sawicki2013}.

Although CDE is the natural minimal generalization to quintessence,
it is still not well explored in the literature. Studies
which consider parametrizations of $w$ are very common, but rarely
consider $c_{s}<1$. As discussed, CDE can play a role in alleviating
the $S_{8}$ tension. From another perspective, there is no indication
that DE can not present relevant fluctuations on small scales. Therefore,
these kinds of models deserve more attention from the cosmology community.

Besides these difficulties, preliminary forecasts studies indicate that future
experiments like Euclid will be able to discriminate between homogeneous and CDE. The
challenges to achieve this goal are big. Most of the constraining power
of Euclid will come from nonlinear scales, and a good understanding
of the nonlinear effects of CDE will be mandatory.

\acknowledgments{This paper is submitted to the Special Issue entitled \textquotedblleft Large 
Scale Structure of the Universe\textquotedblright , led by N. Frusciante
and F. Pace, and belongs to the section \textquotedblleft Cosmology''.
RCB thanks Tiago Castro, Valerio Marra and Franceco Pace for useful
discussions and suggestions.}
\end{paracol}

\bibliographystyle{mdpi.bst}
\bibliography{referencias}

\end{document}